\documentclass[a4paper,11pt]{article}

\usepackage{jheppub}
\usepackage[T1]{fontenc}
\usepackage{hyperref}
\usepackage{graphicx}
\usepackage{amsmath}
\usepackage{epsfig}
\usepackage{subfig}
\usepackage{xcolor}
\usepackage{natbib}

\title{The production of the doubly charmed baryon in deeply inelastic $ep$ scattering at the Large Hadron Electron Collider}

\author[a]{Zhan Sun}
\author{,}
\author[b]{Xing-Gang Wu}
\affiliation[a]{Department of Physics $\&$ Institute of Particle Physics, Guizhou Minzu University, Guiyang 550025, People's Republic of China.}
\affiliation[b]{Department of Physics, Chongqing University, Chongqing 401331, People's Republic of China.}
\emailAdd{zhansun@cqu.edu.cn}
\emailAdd{wuxg@cqu.edu.cn}

\abstract{In this paper, we carry out a detailed study on the production of the doubly charmed baryon in deeply inelastic $ep$ scattering (DIS) for $Q^2\in[2, 100]~\textrm{GeV}^2$, at the Large Hadron Electron Collider (LHeC) with $E_e=60(140)$ GeV and $E_p=7000$ GeV. To exclude the contributions from the diffractive productions and the $b$ hadron decays, we impose the kinematic cuts $0.3<z<0.9$, and $p_{t,\textrm{baryon}}^{\star2}>1~\textrm{GeV}^2$ in the center-of-mass (CM) frame of $\gamma^{*}p$. Based on the designed LHeC luminosity, by detecting the decay channel $\Xi^{+}_{cc}(\Xi^{++}_{cc}) \to \Lambda_c^{+}$ with the subsequent decay $\Lambda_c^{+} \to pK^{-}\pi^{+}$, we predict that about 1880 (2700) $\Xi^{+}_{cc}$ events and 3760 (5400) $\Xi^{++}_{cc}$ events can be accumulated per year, which signifies the prospect of observing them via the DIS at the forthcoming LHeC. In addition, we also predict the distributions of a rich variety of physical observables in the laboratory frame and the $\gamma^{*}p$ CM frame, including $Q^2$, $p_t^{2}$, $Y$(rapidity), $p_t^{\star2}$, $Y^{\star}$, $W$, and $z$ distributions, respectively, which can provide helpful references for studying the doubly charmed baryon. In conclusion, we think that in addition to the LHC, the LHeC is also a helpful platform for studying the properties of the doubly charmed baryon.
}


\begin{document}

\maketitle

\bibliographystyle{JHEP}

\section{Introduction}\label{intro}

The doubly charmed baryon, interpreted as a three-quark state with two heavy $c$ quarks and a light quark $q$ ($q=u,d,s$) based on the quark model \cite{GellMann:1964nj, Zweig, Ebert:1996ec, Gerasyuta:1999pc, Itoh:2000um}, is very helpful for quantitatively testing the quantumn chromodynamics (QCD). It has attracted a great deal of attention from physicists. In year 2002 and 2005, the SELEX collaboration reported the observations of $\Xi^{+}_{cc}$ via its decay channel $\Xi_{cc}^{+} \to \Lambda_{c}^{+}K^{-}\pi^{+}$ \cite{Mattson:2002vu} and $\Xi_{cc}^{+} \to pD^{+}K^{-}$ \cite{Ocherashvili:2004hi}. However, the large production rates released by the SELEX Collaboration have not been confirmed by the BaBAR \cite{Aubert:2006qw}, the Belle \cite{Chistov:2006zj, Kato:2013ynr}, and even the FOCUS \cite{Ratti:2003ez} that is at the same collider of SELEX. In the past several years, the LHCb group has searched for the hadroproduced $\Xi^{+}_{cc}(\Xi^{++}_{cc})$ for many times \cite{Aaij:2013voa, Aaij:2017ueg, Aaij:2018gfl, Aaij:2018wzf,Aaij:2019jfq, Aaij:2019zxa}, collecting fruitful results. In 2013, the LHCb Collaboration performed its first search for $\Xi_{cc}^{+}$, reporting that the measured upper limit of the value of $\mathcal{R}(=\frac{\sigma(\Xi^{+}_{cc})\mathcal{B}(\Xi^{+}_{cc} \to \Lambda_c^{+}K^{-}\pi^{+})}{\sigma(\Lambda_c^{+})})$ is from $1.5 \times 10^{-2}$ (corresponding to the $\Xi_{cc}^{+}$ lifetime of 100 fs) to $3.9 \times 10^{-4}$ (400 fs). While the SELEX observation corresponds to $\mathcal{R}=9\%$ \cite{Mattson:2002vu}. Recently the LHCb group released their latest measurement on the upper limit of $\mathcal{R}$ \cite{Aaij:2019jfq}, varying from $6.5 \times 10^{-3}$ (40 fs) to $0.9 \times 10^{-3}$ (160 fs) for $\sqrt{s}=8$ TeV, and $0.45 \times 10^{-3}$ (40 fs) to $0.12 \times 10^{-3}$ (160 fs) for $\sqrt{s}=13$ TeV, which is still significantly below the value reported by the SELEX Collaboration\footnote{It is important to notice that the production environments of SELEX and LHCb are different, i.e., the collision of the hyperon beam (an admixture of $p,\Sigma^{-}$, and $\pi^{-}$) on a fixed target versus the proton-proton collsion at a much higher energy. Thus some process-dependent effects, e.g., the contributions of the intrinsic or extrinsic charm components in the proton or the hyperon which play a very important role in the low-energy production process \cite{literature21,literature24}, may help to clarify this issue. At this point, perhaps it is still premature to draw a solid conclusion concerning the consistency or inconsistency between the $\mathcal{R}$ values measured by SELEX and LHCb, respectively.}. Regarding $\Xi_{cc}^{++}$, a solid signal was firstly observed by LHCb via the decay channel of $\Xi_{cc}^{++} \to \Lambda_c K^{-}\pi^{+} \pi^{-}$ with $\Lambda_c \to pK^{-}\pi^{+}$ \cite{Aaij:2017ueg} in 2017, which was subsequently confirmed by Refs. \cite{Aaij:2018gfl} and \cite{Aaij:2019zxa}. On theoretical aspect, the doubly charmed baryon productions at various high-energy colliders have been studied in detail in the literatures \cite{literature1, literature2, literature3, literature4, literature5, literature51, literature52, literature6, literature7, literature8, literature9, literature10, literature11, literature12, literature13, literature14, literature15, literature16, literature17, literature18, literature19, literature20, literature201, literature202, literature203, literature21, literature22, literature23, literature24, literature25, literature26, literature27, literature271, literature28, literature29, literature30, literature31}, such as the $e^+e^-$ annihilation, the photoproduction in $ep$ collision, the hadroproductions in $pp$ (or $p\bar{p}$) collision, and the indirect productions via the decays of $Z$ boson, $t$ quark, and the standard Higgs boson. Especially, a dedicated generator GENXICC~\cite{Chang:2007pp, Chang:2009va, Wang:2012vj} has been developed and has been frequently applied to study the production of doubly heavy baryon at the LHC.

In addition to the hadro- and photoproductions, the productions in deeply inelastic scattering (DIS) is another interesting process for the studies of the doubly charmed baryon, like the $J/\psi$ case \cite{Jpsi1, Jpsi2, Jpsi3, Jpsi4, Jpsi5, Jpsi6, Jpsi7, Jpsi8, Jpsi9, Jpsi10, Jpsi101, Jpsi11, Jpsi12, Jpsi13, DISexp1, DISexp2, DISexp3, DISexp4}. Analyzing the productions in DIS at finite $Q^2$ has theoretical and experimental advantages compared to the inelastic photoproduction ($Q^2 \simeq 0$). The large $Q^2$, such as the kinematic region of $Q^2\in[2, 100]~\textrm{GeV}^2$, corresponding to the inclusive $J/\psi$ productions in DIS at the Hadron-Electron Ring Accelerator (HERA), may decrease the theoretical uncertainties \cite{DISexp2} and improve the convergence of the perturbative series~\footnote{Ref.\cite{Jpsi10} shows that the next-to-leading order QCD corrections to the inclusive $J/\psi$ productions in DIS exhibits a rather good convergence, which is much better than the case of hadro- and photoproduction \cite{Jpsihadron1, Jpsihadron2, Jpsihadron3, Jpsiphoto1, Jpsiphoto2, Jpsiphoto3}.}; the contributions via the resolved photon are expected to be negligible since the probability of the resolved photon to emerge decrease rapidly with $Q^{2}$. Furthermore, the background via the diffractive production is also expected to decrease faster with increasing $Q^2$ than the case of the inelastic photoproduction process. On experimental aspects, the distinct signature of the scattered electron in the final state makes the process easier to be detected. Comparing to the hadro- and protoproduction, much more varieties of physical observables can be measured in the DIS process, such as $p_t^{2}$, $Y$(rapidity), $p_t^{\star2}$, $Y^{\star}$, $W$, $z$, and $Q^{2}$ distributions, where the left four variables are related to the doubly charmed baryon.{\em {Throughout the paper we employ the superscript $``\star"$ to denote the measured quantities in the center-of-mass (CM) frame of $\gamma^{*}p$}}.  In view of these advantages, the production in DIS can provide an ideal laboratory for studying the doubly charmed baryon, deserving a separate investigation.

The Large Hadron Electron Collider (LHeC) is designed as a sencond generation of DIS $ep$, and a first electron-ion collider \cite{LHeC1,LHeC2}. As reported in the Conceptual Design Report (CDR) of LHeC \cite{LHeC1}, it takes unique advantage of the intense, high energy hadron beams of the LHC, and a 60 GeV, to possibly 140 GeV, electron beam of high intensity based on a racetrack, energy recovery configuration using two 10 GeV electron linacs. Thus the LHeC will exceed the luminosity of HERA by a factor of $100$, reaching up to $\mathcal{L}=10^{33}\textrm{cm}^{-2}\textrm{s}^{-1}$. Note that this value of $\mathcal{L}$ is based on the original design for LHeC in 2012, from today's experience with the LHC operation, the improved proton beam parameters of the HL-LHC (high luminosity) upgrade may lead to a significantly higher luminosity for the LHeC than the value of $ \mathcal{L}=10^{33}\textrm{cm}^{-2}\textrm{s}^{-1}$. As reported in Ref. \cite{PDG}, by the parasitic operation in parallel to the HL-LHC $pp$ collision, the up-to-date luminosity of LHeC could be improved to be $ \mathcal{L}=0.8 \times 10^{34}\textrm{cm}^{-2}\textrm{s}^{-1}$, about one order of magnitude larger than the designed value in 2012. From these perspectives, the forthcoming LHeC allows a multitude of crucial DIS measurements to be performed. Thus, in this paper, we will for the first time carry out the studies on the doubly charmed baryon productions in DIS ($ep$) at the LHeC.

The rest of the paper is organized as follows: In section \ref{intro}, we give a description on the calculation formalism. In section \ref{cal}, the phenomenological results and discussions are presented. Section \ref{sum} is reserved as a summary.

\section{Calculation Formalism}\label{cal}

\subsection{General Formalism}

It is a widely accepted view that the productions of the doubly charmed baryon can be factorized into two steps, c.f. Refs.\cite{literature7, literature23, literature24, literature29}:
\begin{itemize}
\item[1)]
The production of a $c$ quark pair by the perturbative calculable hard processes with the subsequent $c$ quark pair nonperturbatively forming a bounding $(cc)$-diquark with color- and spin- configuration $[n]$. According to the decomposition $3 \otimes 3=\bar{\textbf{3}} \oplus \textbf{6}$ in $\textrm{SU}_{\textrm{c}}(3)$ group, the diquark can only be in $\bar{\textbf{3}}$ and $\textbf{6}$ color state. Based on nonrelativistic Quantum Chromodynamics (NRQCD) \cite{NRQCD}, at the leading order of $v_c$ (the relative velocity of the two constituent $c$ quarks in the diquark), there are only two $S-$wave configurations for the case of $(cc)-$diquark \footnote{Other diquark configurations, i.e., $(cc)[^1S_0]_{\bar{\textbf{3}}}$ and $(cc)[^3S_1]_{\textbf{6}}$ are forbidden due to the Fermi-Dirac statistics for identical particles.} \cite{literature7, literature23, literature24}, i.e., the spin-triplet $n=[^3S_1]_{\bar{\textbf{3}}}$ and spin-singlet $n=[^1S_0]_{\textbf{6}}$ that can be described by a matrix element, $h_{\bar{\textbf{3}}(\textbf{6})}$. Regarding the $P-$wave processes, the contributions shall be at least $v_c^2-$suppressed to that of the $S-$wave processes. Taking the $\Xi_{cc}$ hadroproductions as an example, the $P-$wave contributions just account for about $3\%-5\%$ of the total result \cite{Berezhnoy:2020aox}. In view of these points, in this paper we just concentrate on the dominant $S-$wave contributions of $(cc)[^3S_1]_{\bar{\textbf{3}}}$ and $(cc)[^1S_0]_{\textbf{6}}$.
\item[2)]
The hadronization of the diquark into a physical colorless baryon $\Xi_{ccq}$ ($q=u,d,s$) by grabbing a light quark with possible soft gluons from the hadron.  {\em For convenience, hereinafter we label the doubly charmed baryon as $\Xi_{cc}$ instead of $\Xi_{ccq}$}. During the hadronization procedure of the diquark into $\Xi_{cc}$, one usually assumes the total evolving probability to be $100\%$, referring as the $``\textrm{direct evolution}"$. Among this total $``100 \%"$ probability, the diquark fragmenting into $\Xi^{+}_{cc}(ccd)$ and $\Xi^{++}_{cc}(ccu)$ both accounts for $43\%$, respectively, and the ratio for $\Omega^{+}_{cc}(ccs)$ is $14\%$ \cite{literature27,Sjostrand:2006za}. Chen et al. in Ref. \cite{literature16} pointed out that the direct evolution and the evolution via the heavy quark fragmentation model suggested by Ref. \cite{fragmentation} lead to almost the same numerical results, indicating that the direct evolution mechanism is precise enough to describe the production of the doubly charmed baryon.
\end{itemize}

\begin{figure}[!h]
\begin{center}
\hspace{0cm}\includegraphics[width=0.65\textwidth]{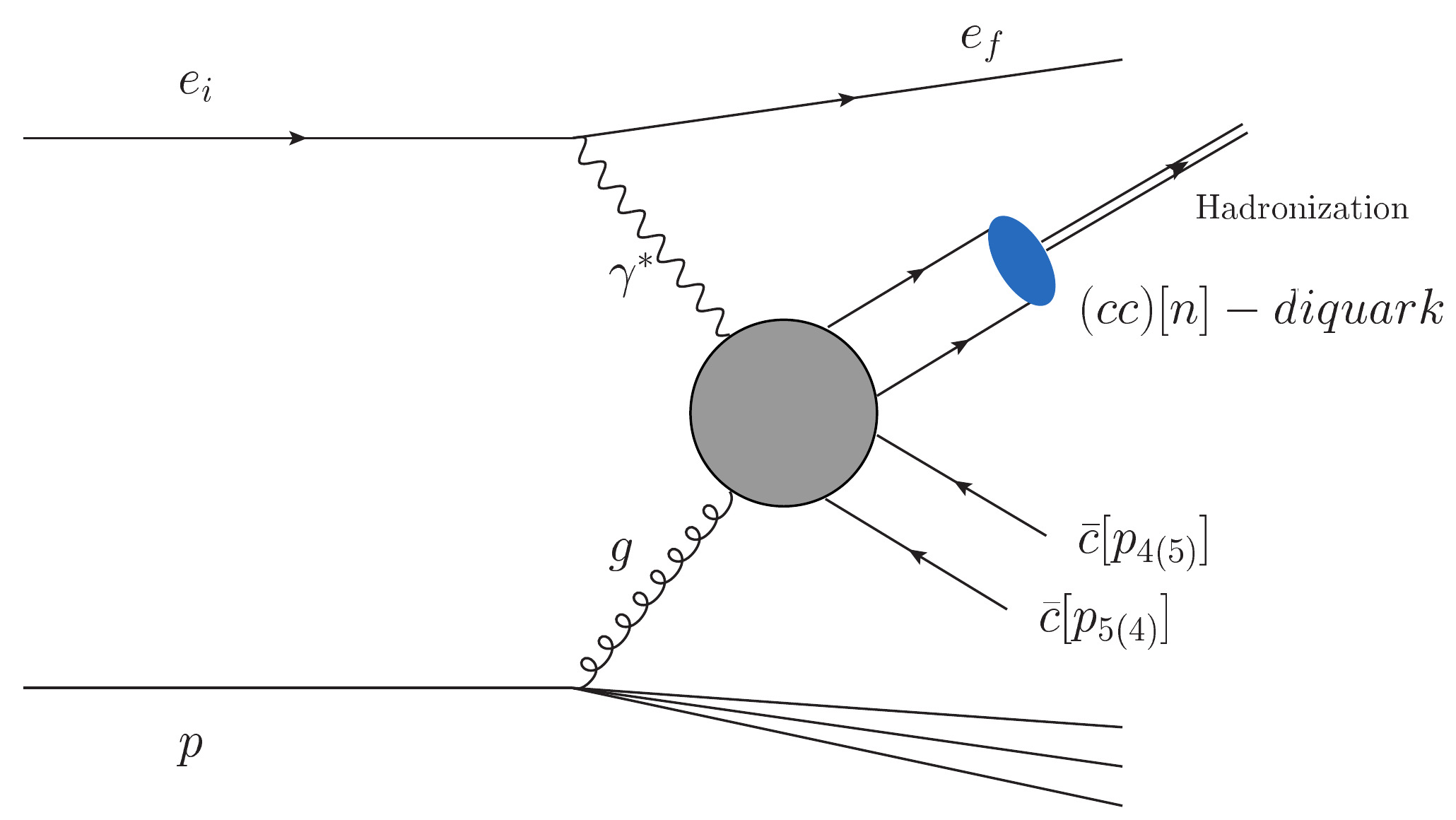}
\caption{\label{fig:Feyn}
The illustrative diagram for the $\Xi_{cc}$ production in deeply inelastic $ep$ scattering.}
\end{center}
\end{figure}

According to the above steps, we draw the illustrative diagram regarding the DIS production process $e+p \to e+\Xi_{cc}+X$ in Figure \ref{fig:Feyn}, where $e_i$, $e_f$, $\gamma^{*}$, $g$, and $p$ represent the initial and final electron, the virtual photon, the gluon parton, and the proton, respectively. To be specific, we shall focus on the contributions from the dominant gluon partonic process $e+g \to e+\Xi_{cc}+\bar{c}+\bar{c}$; and the less-important contributions from the $c-$quark content in proton shall not be considered here. In addition, the observed value of $Q$ is smaller than the mass of the $Z$ boson, such as $Q^2\in[2, 100]~{GeV}^2$, thus we neglect the contributions from the $Z$ propagator, as well as the $H^{0}$ propagator, taking only $\gamma^{*}$ into consideration. By regarding the electron and proton as massless, we introduce some generally used invariants to describe the DIS process
\begin{eqnarray}
&&Q^2=-p_{\gamma^{*}}^2=-(p_{e_{i}}-p_{e_{f}})^2, \nonumber \\
&&W^2=(p_{\gamma^*}+p_p)^2, \nonumber \\
&&S=(p_p+p_{e_{i}})^2=2~p_{p} \cdot p_{e_{i}}, \nonumber \\
&&\hat{s}=(p_g+p_{\gamma^{*}})^2, \nonumber \\
&&s=\hat{s}+Q^2=2~p_{g} \cdot p_{\gamma^{*}},\nonumber \\
&&z=\frac{p_{p} \cdot p_{\Xi_{cc}}}{p_{p} \cdot p_{\gamma^{*}}},~~~y=\frac{p_p \cdot p_{\gamma^{*}}}{p_p \cdot p_{e_i}}.
\end{eqnarray}
Based on the NRQCD and the collinear factorization, we factorize the $\Xi_{cc}$ production in DIS via $e+p \to e+\Xi_{cc}+\bar{c}+\bar{c}$ as
\begin{eqnarray}
&&d\sigma(e+p \to e+\Xi_{cc}+\bar{c}+\bar{c})= \nonumber \\
&&=\int~dx~\sum_{n} f_{g/p}(x,\mu_{f}) d\hat{\sigma}_{e+g \to e+(cc)[n]+\bar{c}+\bar{c}} \times \langle\mathcal O^{\Xi_{cc}}(n)\rangle,
\label{cross section}
\end{eqnarray}
where $d\hat{\sigma}_{e+g \to e+(cc)[n]+\bar{c}+\bar{c}}$ is the perturbative calculable short distance coefficient (SDC), representing the production of a configuration of the $(cc)[n]$ intermediate quark pair with $n=[^3S_1]_{\bar{\textbf{3}}}$ or $[^1S_0]_{\textbf{6}}$. The universal nonperturbative long distance matrix element $\langle\mathcal O^{\Xi_{cc}}(n)\rangle$ stands for the probability of $(cc)[n]$ pair into the corresponding diquark, subsequently fragmenting into $\Xi_{cc}$, i.e., $h_{\bar{\textbf{3}}(\textbf{6})} \times 100\%$ based on the direct evolution mechanism. $f_{g/p}(x,\mu_{f})$ is the gluon parton distribution function evaluated at the factorization scale $\mu_f$. The hard partonic SDC can be written as
\begin{eqnarray}
d\hat{\sigma}_{e+g \to e+(cc)[n]+\bar{c}+\bar{c}}=\frac{1}{2 xS}\frac{1}{N_c N_s} |\mathcal{M}|^2 d\Phi.
\end{eqnarray}
where $1/(N_c N_s)$ is the colour and spin average factor; $|\mathcal{M}|^2$ and $d\Phi$ are squared matrix element and the 4-body phase space, respectively. In the following subsections, we depict the method of how to calculate $|\mathcal{M}|^2$ and $d\Phi$.

\subsection{$|\mathcal{M}|^2$}\label{M2}

\begin{figure}[!h]
\begin{center}
\hspace{0cm}\includegraphics[width=1.0\textwidth]{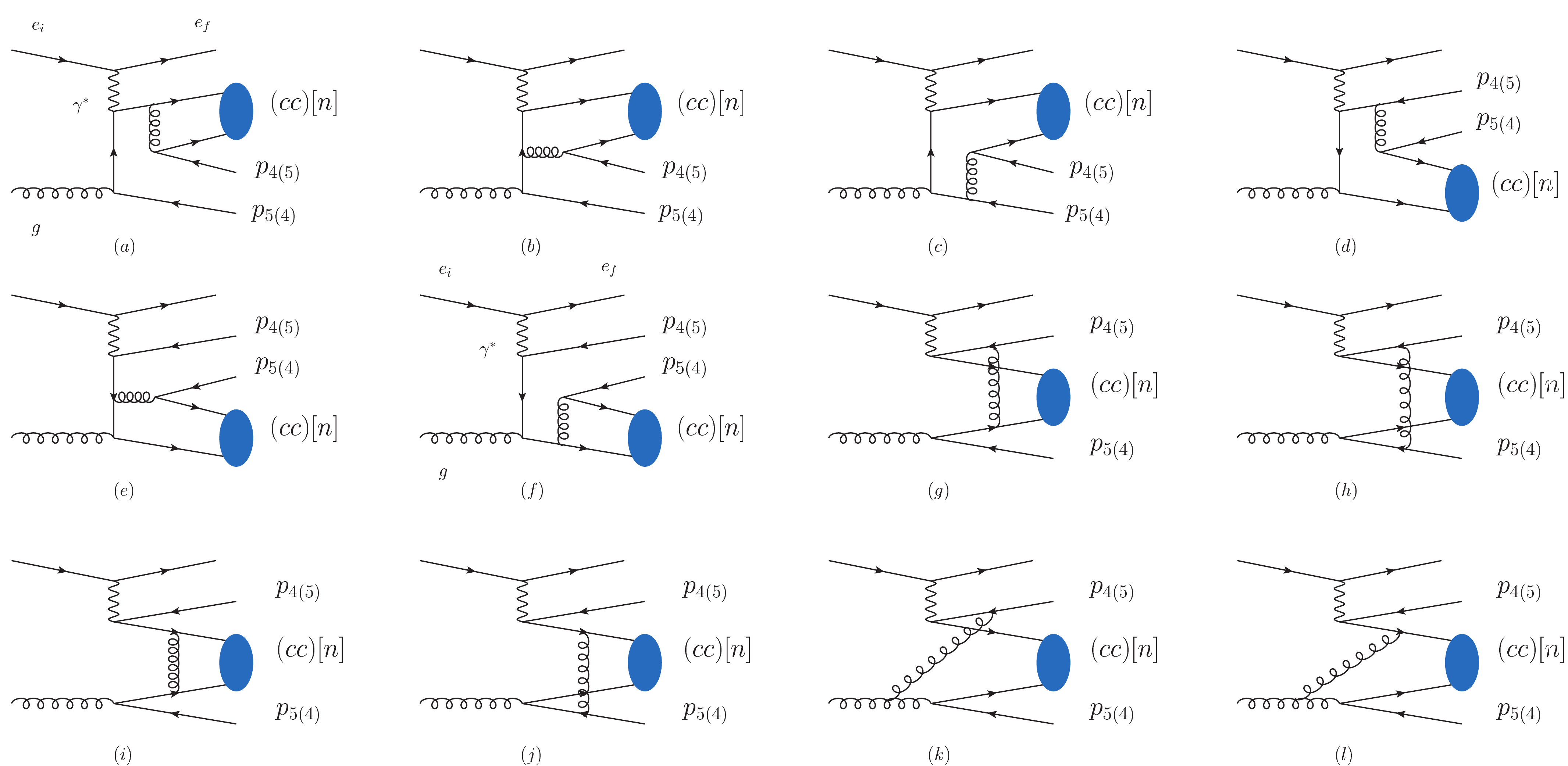}
\caption{\label{fig:Feyn1}
The Feynman diagrams for the subprocesses $e+g \to e+(cc)[n]+\bar{c}+\bar{c}$ ($p_4 \leftrightarrow p_5$).}
\end{center}
\end{figure}

For the subprocess $e+g \to e+(cc)[n]+\bar{c}+\bar{c}$, there are in total 48 Feynman diagrams, half of which are illustrated in Figure \ref{fig:Feyn1} ($p_4 \leftrightarrow p_5$). The other 24 ones can be obtained by exchanging the two identical $c-$quark lines inside the diquark. Seeing that we have set the relative velocity between the two constituent $c$ quarks in the diquark to be zero, the two parts of 24 diagrams contribute identically. Thus at the cross section level we only need to calculate the 24 diagrams in Figure \ref{fig:Feyn1}, and multiply a factor of $2^2$. Simultaneously we should introduce an additional factor of $1/(2!2!)$ to deal with the identity of the two constituent $c$ quarks inside the diquark, and that of the two $\bar{c}$ quarks in the final states. To calculate $|\mathcal{M}|^2$, we decompose it as
\begin{eqnarray}
|\mathcal{M}|^2=L^{\mu\nu}H_{\mu\nu},
\end{eqnarray}
where $L^{\mu\nu}$ denote the leptonic tensor regarding $e_i \to e_f+\gamma^{*}$; $H_{\mu\nu}$ represents the hadronic tensor corresponding to $\gamma^{*}+g \to (cc)[n]+\bar{c}+\bar{c}$.

\subsubsection{Leptonic tensor}

By a direct calculation, one can obtain
\begin{eqnarray}
L^{\mu\nu}&=&\frac{1}{(Q^2)^2}4 \pi \alpha \textrm{Tr}[p\!\!\!\slash_{e_i} \gamma_{\mu} p\!\!\!\slash_{e_f} \gamma_{\nu}]\nonumber \\
&=&\frac{1}{(Q^2)^2}8 \pi \alpha Q^2\left( -g^{\mu\nu}+\frac{4p_{e_i}^{\mu}p_{e_i}^{\nu}-2p_{e_i}^{\mu}p_{\gamma^{*}}^{\nu}-2p_{\gamma^{*}}^{\mu}p_{e_i}^{\nu}}{Q^2} \right)\nonumber \\
&=&\frac{8 \pi \alpha}{Q^2} l^{\mu\nu}.
\label{leptonic tensor}
\end{eqnarray}
If only one final-state hadron (e.g. $\Xi_{cc}$) other than the scatterd electron is observed, $l^{\mu\nu}$ can be decomposed into the linear combination of four independent Lorentz invariant structures as \cite{Jpsi8,Jpsi9,Jpsi10}
\begin{eqnarray}
l^{\mu\nu}=A_g \left(-g^{\mu\nu}-\frac{p^{\mu}_{\gamma^{*}}p^{\nu}_{\gamma^{*}}}{Q^2}\right)+A_L
\epsilon^{\mu}_L \epsilon^{\nu}_L+A_{LT}(\epsilon^{\mu}_L \epsilon^{\nu}_T+\epsilon^{\mu}_T \epsilon^{\nu}_L)+A_T\epsilon^{\mu}_T \epsilon^{\nu}_T
\label{leptonic tensor norm}
\end{eqnarray}
where
\begin{eqnarray}
&&\epsilon_{L}=\frac{1}{Q}\left(p_{\gamma^{*}}+\frac{2Q^2 }{s} p_g \right), \nonumber \\
&&\epsilon_{T}=\frac{1}{p_t^{\star}}\left(p_{\Xi_{cc}}-\rho p_{g}-zp_{\gamma^{*}} \right),
\label{epsilon}
\end{eqnarray}
and
\begin{eqnarray}
&&A_g=1+\frac{2(1-y)}{y^2}-\frac{2(1-y)}{y^2}\cos(2\psi^{\star}), \nonumber \\
&&A_L=1+\frac{6(1-y)}{y^2}-\frac{2(1-y)}{y^2}\cos(2\psi^{\star}), \nonumber \\
&&A_{LT}=\frac{2(2-y)}{y^2}\sqrt{1-y}\cos(\psi^{\star}), \nonumber \\
&&A_{T}=\frac{4(1-y)}{y^2}\cos(2\psi^{\star}),
\end{eqnarray}
in which $\rho$ is defined as
\begin{eqnarray}
\rho=\frac{\left(p^{\star2}_t+M^2_{\Xi_{cc}}\right)/z+zQ^2}{s}.
\end{eqnarray}
Here, $p^{\star2}_t$ is the square of the transverse momentum of $\Xi_{cc}$; $\psi^{\star}$ denotes the azimuthal angle of the hadron ($\Xi_{cc}$) production plane around the $z$ axis relative to the lepton plane expanded by the incoming and the outgoing electrons. Substituting Equation (\ref{epsilon}) into Equation (\ref{leptonic tensor norm}), with the help of the current conservation equation $p^{\mu}_{\gamma^{*}}H_{\mu\nu}=0$, one can obtain a more convenient form of $l^{\mu\nu}$
\begin{eqnarray}
l^{\mu\nu}=C_1(-g^{\mu\nu})+C_2 p^{\mu}_g p^{\nu}_g + C_3 \frac{p^{\mu}_g p^{\nu}_{\Xi_{cc}}+p^{\mu}_{\Xi_{cc}} p^{\nu}_{g}}{2}+C_4 p^{\mu}_{\Xi_{cc}} p^{\nu}_{\Xi_{cc}},
\label{leptonic tensor new}
\end{eqnarray}
where
\begin{eqnarray}
C_1&=&A_g, \nonumber \\
C_2&=&\frac{4Q^2}{s^2}(A_L-2 \beta A_{LT}+\beta^2A_T), \nonumber \\
C_3&=&\frac{4Q}{p^{\star}_t s}(A_{LT}-\beta A_T),\nonumber \\
C_4&=&\frac{1}{p_t^{\star2}}A_T,
\end{eqnarray}
with
\begin{eqnarray}
\beta=\frac{\left(p^{\star2}_t+M^2_{\Xi_{cc}}\right)/z+zQ^2}{2p^{\star}_t Q}.
\end{eqnarray}
Comparing the two leptonic tensor forms in Equations (\ref{leptonic tensor}) and (\ref{leptonic tensor new}), one can find that the four basic tensors in Equation (\ref{leptonic tensor new}) are only correlated to the hadronic process momentum ($p_g,p_{\Xi_{cc}}$), which can greatly reduce the complication of the subsequent contraction with $H_{\mu\nu}$. In Refs.\cite{Jpsi9, Jpsi10}, the leptonic tensor in Equation (\ref{leptonic tensor new}) have been utilized to calculate the $J/\psi$ productions in DIS at HERA.

\subsubsection{Hadronic tensor}

As stated in section \ref{M2}, there are 48 Feynman diagrams for the subprocess $e+g \to e+(cc)[n]+\bar{c}+\bar{c}$, which are representatively illustrated in Fig. \ref{fig:Feyn1}. Because the calculation procedures for all the diagrams in Fig. \ref{fig:Feyn1} are similar, without loss of generality, in the following we just take the first diagram (Fig. \ref{fig:Feyn1}(a)) as an example to describe how to deal with the hadronic amplitude. By assuming that the virtual photon is attached to the fermion line including $v(p_5)$, we can write the hadronic amplitude of Fig. \ref{fig:Feyn1}(a) as
\begin{eqnarray}
\mathcal{M}^{H,1}_{\mu,\sigma}&=&\mathcal{C} g_s^3 e_c e \frac{1}{(p_4+p_{\Xi_{cc}})^2-m_c^2}\frac{1}{(p_g-p_5)^2-m_c^2} \frac{1}{(p_4+\frac{p_{\Xi_{cc}}}{2})^2} \nonumber \\
& \times & \left\{\bar{u} \left(\frac{p_{\Xi_{cc}}}{2}\right) \Gamma_{\rho}  s_f(p_4+p_{\Xi_{cc}},m_c) \Gamma_{\mu} s_f(p_g-p_5,m_c) \Gamma_{\sigma} v(p_5)\right\} \nonumber \\
& \times & \left\{\bar{u}\left(\frac{p_{\Xi_{cc}}}{2}\right) \Gamma_{\rho} v(p_4)\right\},
\label{hadronic amplitude}
\end{eqnarray}
where $s_f(k,m)=k\!\!\!\slash+m$ is the fermion propagator, $\Gamma_{\mu,\rho,\sigma}$ are the interaction vertices, $\mathcal{C}$ is the color factor, and $e_c$ is the electric charge of $c$ quark. For the first fermion line in Equation (\ref{hadronic amplitude}),
\begin{eqnarray}
\mathcal{A}=\bar{u} \left(\frac{p_{\Xi_{cc}}}{2}\right) \Gamma_{\rho}  s_f(p_4+p_{\Xi_{cc}},m_c) \Gamma_{\mu} s_f(p_g-p_5,m_c) \Gamma_{\sigma} v(p_5),
\label{hadronic amplitude1}
\end{eqnarray}
we have
\begin{eqnarray}
\mathcal{A}&=&\mathcal{A}^{T} \nonumber \\
&=&v^{T}(p_5) \Gamma^{T}_{\sigma} s^{T}_f(p_g-p_5,m_c) \Gamma^{T}_{\mu} s^{T}_f(p_4+p_{\Xi_{cc}},m_c) \Gamma^{T}_{\rho} \bar{u}
\left(\frac{p_{\Xi_{cc}}}{2}\right)^{T}.
\end{eqnarray}
By inserting the charge conjugate matrix $C=-i \gamma^{2} \gamma^{0}$, which satisfies the following equations
\begin{eqnarray}
&&CC^{-1}=1,~~~v^{T}(p_5)C=-\bar{u}(p_5),~~~C^{-}\bar{u}\left(\frac{p_{\Xi_{cc}}}{2}\right)^{T}
=v\left(\frac{p_{\Xi_{cc}}}{2}\right), \nonumber \\
&&C^{-} \Gamma^{T}_{\mu,\rho,\sigma} C=-\Gamma_{\mu,\rho,\sigma},~~~C^{-}s^{T}_f(k,m_c)C=s_f(-k,m_c),
\end{eqnarray}
we obtain
\begin{eqnarray}
\mathcal{A}&=&v^{T}(p_5) C C^{-}\Gamma^{T}_{\sigma} C C^{-} s^{T}_f(p_g-p_5,m_c) C C^{-} \Gamma^{T}_{\mu}C C^{-} s^{T}_f(p_4+p_{\Xi_{cc}},m_c) \nonumber \\
&\times & C C^{-} \Gamma^{T}_{\rho} C C^{-} \bar{u} \left(\frac{p_{\Xi_{cc}}}{2}\right)^{T} \nonumber \\
&=&(-1)^{\xi+1} \bar{u}(p_5) \Gamma_{\sigma} s_f(p_5-p_g,m_c) \Gamma_{\mu} s_f(-p_4-p_{\Xi_{cc}},m_c) \Gamma_{\rho} v\left(\frac{p_{\Xi_{cc}}}{2}\right),
\label{hadronic amplitude11}
\end{eqnarray}
where $\xi(=3)$ is the number of the interaction vertices in $\mathcal{A}$.
Combining Equations (\ref{hadronic amplitude}), (\ref{hadronic amplitude1}), and (\ref{hadronic amplitude11}), the product of the two fermion lines in $\mathcal{M}^{H,1}_{\mu,\sigma}$ can be transformed as
\begin{eqnarray}
&&(-1)^{\xi+1}\bar{u}(p_5)  \Gamma_{\sigma} s_f(p_5-p_g,m_c) \Gamma_{\mu} s_f(-p_4-p_{\Xi_{cc}},m_c) \Gamma_{\rho} \nonumber \\
&&\times v \left(\frac{p_{\Xi_{cc}}}{2}\right) \bar{u}\left(\frac{p_{\Xi_{cc}}}{2}\right) \Gamma_{\rho} v(p_4) \nonumber \\
& = & (-1)^{\xi+1}\bar{u}(p_5)  \Gamma_{\sigma} s_f(p_5-p_g,m_c) \Gamma_{\mu} s_f(-p_4-p_{\Xi_{cc}},m_c) \Gamma_{\rho} \nonumber \\
&&\times \Pi^{0(\zeta)}_{p_{\Xi_{cc}}} \times \Gamma_{\rho} v(p_4),
\label{hadronic amplitude2}
\end{eqnarray}
where $\Pi^{0}$ and $\Pi^{\zeta}$ are the spin projector operators \cite{projector}
\begin{eqnarray}
&&\Pi^{0}_q=\frac{1}{\sqrt{8m_c^3}} \left(\frac{q\!\!\!\slash}{2}-m_c\right) \gamma^{5} \left(\frac{q\!\!\!\slash}{2}+m_c\right), \nonumber \\
&&\Pi^{\zeta}_q=\frac{1}{\sqrt{8m_c^3}} \left(\frac{q\!\!\!\slash}{2}-m_c\right) \gamma^{\zeta} \left(\frac{q\!\!\!\slash}{2}+m_c\right),
\end{eqnarray}
which are for the spin singlet $(^1S_0)$ and the spin triplet $(^3S_1)$, respectively. Finally, we have
\begin{eqnarray}
\mathcal{M}^{H,1}_{\mu,\sigma}&=&\mathcal{C} g_s^3 e_c e \frac{1}{(p_4+p_{\Xi_{cc}})^2-m_c^2}\frac{1}{(p_g-p_5)^2-m_c^2} \frac{1}{(p_4+\frac{p_{\Xi_{cc}}}{2})^2} \nonumber \\
& \times & (-1)^{\xi+1}\bar{u}(p_5)  \Gamma_{\sigma} s_f(p_5-p_g,m_c) \Gamma_{\mu} s_f(-p_4-p_{\Xi_{cc}},m_c) \Gamma_{\rho} \nonumber \\
& \times & \Pi^{0(\zeta)}_{p_{\Xi_{cc}}} \times \Gamma_{\rho} v(p_4).
\label{hadronic amplitude new}
\end{eqnarray}
From Equation (\ref{hadronic amplitude new}), it is not difficult to find the calculation for the doubly charmed baryon is similar to the charmonium case, except for the factor of $(-1)^{\xi+1}$ and the color factor.

The color factor in Equation (\ref{hadronic amplitude new}), corresponding to Fig. \ref{fig:Feyn1}(a), can be written as
\begin{equation}
\mathcal{C}=T^{a}_{il} T^{b}_{ln} \times T^{a}_{jm}
\label{colorini}
\end{equation}
where $b=1 \cdots 8$ is the color indices of the incoming gluon. $i,j=1,2,3$ and $m,n=1,2,3$ denote the color indices of the two constituent $c$ quarks in the diquark and that of the two final-state $\bar{c}$ quarks, respectively. By the fact that $3 \otimes 3=\bar{\textbf{3}} \oplus \textbf{6}$ in $\textrm{SU}_{\textrm{c}}(3)$ group, the diquark can be either in anti-triplet $\bar{\textbf{3}}$ or in sextuplet $\textbf{6}$ color state. Based on this, we introduce the function $\frac{G_{ijk}}{\sqrt{2}}$ to describe the diquark color with $\sqrt{2}$ denoting the normalized factor, where $k=3$ represents the color indices of the diquark. $G_{ijk}$ is identical to the antisymmetric $\varepsilon_{ijk}$ ($\bar{\textbf{3}}$) and the symmetric $f_{ijk}$ ($\textbf{6}$),
which satisfies the following equations
\begin{eqnarray}
\varepsilon_{ijk}\varepsilon_{i^{'}j^{'}k}&=&\delta_{i i^{'}}\delta_{j j^{'}}-\delta_{j i^{'}}\delta_{i j^{'}}, \nonumber \\
f_{ijk}f_{i^{'}j^{'}k}&=&\delta_{i i^{'}}\delta_{j j^{'}}+\delta_{j i^{'}}\delta_{i j^{'}}.
\label{color}
\end{eqnarray}
Then $\mathcal{C}$ in Equation (\ref{colorini}) can be rewritten as 
\begin{eqnarray}
\mathcal{C} &=& T^{a}_{il} T^{b}_{ln} \times T^{a}_{jm}\times\frac{G_{ijk}}{\sqrt{2}}\sim \mathcal{C}_{B},
\label{color1}
\end{eqnarray}
where $\mathcal{C}_{B}$ is identical to $T^{b}_{ln} \times G_{mlk}$. Note that, the color factors of the diagrams in Fig. \ref{fig:Feyn1} can all be reduced into a product of $\mathcal{C}_{B}$ and some constants. With the help of the two relations in Equation (\ref{color}), $|\mathcal{C}_B|^2=8$ for anti-triplet $\bar{\textbf{3}}$, and 16 for sextuplet $\textbf{6}$.

Combining Equations (\ref{hadronic amplitude new}) and (\ref{color1}), one can obtain the complete form of $\mathcal{M}^{H,1}_{\mu,\sigma}$. In a similar way, the amplitudes for the other diagrams in Fig. \ref{fig:Feyn1} can be directly achieved. By summing all the amplitudes together and summing over the polarization vector of the incoming gluon, we finally obtain the hadronic tensor $H_{\mu\nu}$.

\subsection{Phase space}

To deal with the 4-body phase space $d\Phi$, we first decompose it into two components
\begin{eqnarray}
d\Phi&=&d\Phi_{L}d\Phi_{H},
\end{eqnarray}
where
\begin{eqnarray}
d\Phi_{L}&\equiv&\frac{d^3 p_{e_f}}{(2\pi)^3 2p^0_{e_f}}, \nonumber \\
d\Phi_{H}&\equiv&(2\pi)^4\delta^4(p_{\gamma^{*}}+p_g-p_{\Xi_{cc}}-p_{4}-p_{5})\frac{d^3 p_{\Xi_{cc}}}{(2\pi)^3 2p^0_{\Xi_{cc}}}\frac{d^3 p_{4}}{(2\pi)^3 2p^0_{4}}\frac{d^3 p_{5}}{(2\pi)^3 2p^0_{5}}.\nonumber \\
\end{eqnarray}
Integrating over the azimuthal angle of the outgoing electron, one can obtain
\begin{eqnarray}
d\Phi_{L}=\frac{1}{(4\pi)^2 S} dQ^2 dW^2.
\label{phiL}
\end{eqnarray}
By introducing a new momentum $p_{45}=p_4+p_5$ with $p^2_{45}=s_1$, $d\Phi_{H}$ can be rewritten as
\begin{eqnarray}
d\Phi_{H}&=&\left[(2\pi)^4\delta^4(p_{\gamma^{*}}+p_g-p_{\Xi_{cc}}-p_{45})\frac{d^3 p_{\Xi_{cc}}}{(2\pi)^3 2p^0_{\Xi_{cc}}}\frac{d^3 p_{45}}{(2\pi)^3 2p^0_{45}}\right] \nonumber \\
&\times& \left[(2\pi)^4\delta^4(p_{45}-p_4-p_5)\frac{ds_1}{2\pi}\frac{d^3 p_{4}}{(2\pi)^3 2p^0_{4}}\frac{d^3 p_{5}}{(2\pi)^3 2p^0_{5}}\right] \nonumber \\
&=&d\Phi_{H_1} \times d\Phi_{H_2}. \label{phase space}
\end{eqnarray}
Integrating over the three-momentum of $p_{45}$, we have (see e.g. Ref. \cite{Jpsi9})
\begin{eqnarray}
dxd\Phi_{H_1}&=&\frac{1}{32 \pi^2 p^{0}_{45}}\delta(p^0_{\gamma^{*}}+p^0_g-p^0_{\Xi_{cc}}-p^0_{45})dp^{\star2}_{t}\frac{dz}{z} d\psi^{\star} dx \nonumber \\
&=&\frac{1}{(4\pi)^2 (W^2+Q^2)z(1-z)} dp^{\star2}_{t} dz d\psi^{\star},
\label{phiH1}
\end{eqnarray}
In the second step of Equation (\ref{phiH1}), we have utilized $\delta(p^0_{\gamma^{*}}+p^0_g-p^0_{\Xi_{cc}}-p^0_{45})$ to integrate over $dx$, thus the value of $x$ has been fixed at
\begin{eqnarray}
x=\frac{a+Q^2}{W^2+Q^2},~\textrm{with}~a=\frac{p^{\star2}_t+M^2_{\Xi_{cc}}}{z}+\frac{p^{\star2}_t+s_1}{1-z}.
\end{eqnarray}

As for $d\Phi_{H_2}$, in the CM frame of the two final-state $\bar{c}$, we have
\begin{eqnarray}
&&p_{45}=(\sqrt{s_1},0,0,0), \nonumber \\
&&p_{4}=(p^0_4,|\overrightarrow{p_4}|\sin{\theta_{\bar{c}}} \cos{\phi_{\bar{c}}},|\overrightarrow{p_4}|\sin{\theta_{\bar{c}}} \sin{\phi_{\bar{c}}},|\overrightarrow{p_4}|\cos{\theta_{\bar{c}}}), \nonumber \\
&&p_{5}=(p^0_5,-|\overrightarrow{p_4}|\sin{\theta_{\bar{c}}} \cos{\phi_{\bar{c}}},-|\overrightarrow{p_4}|\sin{\theta_{\bar{c}}} \sin{\phi_{\bar{c}}},-|\overrightarrow{p_4}|\cos{\theta_{\bar{c}}}).
\end{eqnarray}
Then
\begin{eqnarray}
d\Phi_{H_2}&=&(2\pi)^4\delta^4(p_{45}-p_4-p_5)\frac{ds_1}{2\pi}\frac{d^3 p_{4}}{(2\pi)^3 2p^0_{4}}\frac{d^3 p_{5}}{(2\pi)^3 2p^0_{5}} \nonumber \\
&=&\frac{d s_1}{2\pi}\frac{1}{(2\pi)^2}\frac{|\overrightarrow{p_4}|}{2}dp^0_4 d\Omega_{\bar{c}} d^4p_5 \delta(p^2_5-m^2_{\bar{c}}) \delta^4(p_{45}-p_4-p_5) \nonumber \\
&=&\frac{1}{(4\pi)^3}\sqrt{1-\frac{4m_{\bar{c}}^2}{s_1}} ds_1 d\cos{\theta_{\bar{c}}} d\phi_{\bar{c}}.
\label{phiH2}
\end{eqnarray}
Combining Equations (\ref{phiL}), (\ref{phiH1}), and (\ref{phiH2}), we finally obtain\footnote{Note that, the derivation of $d\Phi_{H_2}$ is based on the CM frame of $p_4 p_5$, thus in the calculations one should transform to the $\gamma^{*}p$ CM frame, or the laboratory frame.}
\begin{eqnarray}
dxd\Phi=\frac{1}{(4\pi)^7 S (W^2+Q^2) z (1-z)}\sqrt{1-\frac{4m_{\bar{c}}^2}{s_1}} dQ^2 dW^2 dp^{\star2}_t dz d\psi^{\star} ds_1 d\cos{\theta_{\bar{c}}} d\phi_{\bar{c}}. \nonumber \\
\end{eqnarray}

In our calculations, the mathematica-fortran package $\textrm{M}\scriptsize{\textrm{ALT@FDC}}$ that has been employed to calculate several heavy quarkonium related DIS processes \cite{Jpsi8,Jpsi9,Jpsi10} is used to deal with $H_{\mu\nu}$ and $|\mathcal{M}|^2(=L^{\mu\nu}H_{\mu\nu})$, as well as the phase space integration.

\section{Phenomenological results}\label{results}

\subsection{Kinematic cuts and input parameters}

According to the CDR of LHeC \cite{LHeC1, LHeC2}, the designed beam energy of the incident electron and proton are: $E_e=60$ GeV, to possibly 140 GeV, and $E_P=7000$ GeV. Reviewing the inclusive $J/\psi$ productions in DIS at HERA \cite{DISexp1, DISexp2, DISexp3, DISexp4}, the measurements mainly cover the range of $p_{t,J/\psi}^{\star2}>1~\textrm{GeV}^2$, $0.3<z<0.9$, $2<Q^2<100~\textrm{GeV}^2$, and $50<W<225$ GeV. The cuts regarding $p_{t,J/\psi}^{\star2}$ and $z$ are applied to suppress the effects via the diffractive productions, which cannot yet be reliably described by purely perturbative QCD calculations, and that from the $b$ hadrons decay. As for the $\Xi_{cc}$ case, the situation is similar, so we shall also carry out the cut operation to avoid kinematic overlaps with the diffractive productions and the decay of $b$ hadrons. Considering the HERA is the unique available $ep$ collider by now, it is natural and reasonable to take its experimental conditions for reference to assume the kinematic cuts for the $\Xi_{cc}$ productions in DIS at LHeC. To be specific, in our calculations, we adopt the following cut conditions
\begin{eqnarray}
&&p_{t,\Xi_{cc}}^{\star2}>1~\textrm{GeV}^2,~W>50~\textrm{GeV}, \nonumber \\
&&2<Q^2<100~\textrm{GeV}^2, \nonumber \\
&&0.3<z<0.9.
\label{cuts}
\end{eqnarray}
The other input parameters are listed below
\begin{itemize}
\item[1)]
The charm quark mass is taken as $m_c=M_{\Xi_{cc}}/2=1.75$ GeV \cite{literature24, literature27}; the fine structure constant is $\alpha=1/128.$
\item[2)]
The default factorization and renormalization scales are chosen to be $\mu_f=\mu_r=\mu_0=\sqrt{Q^2+M^2_{\Xi_{cc}}}$.
\item[3)]
According to the velocity scaling rule of NRQCD, we take the usual assumption that $h_{\bar{\textbf{3}}}$ and $h_{\textbf{6}}$ have the equal values \cite{literature7, literature23, literature24}, which are related to the wave function at the origin \cite{literature3, literature23}
\begin{eqnarray}
h_{\bar{\textbf{3}}}=h_{\textbf{6}}=|\Psi_{(cc)}(0)|^2=0.039~\textrm{GeV}^3.
\end{eqnarray}
\end{itemize}

\subsection{Integrated cross sections}

The integrated cross sections of $\Xi_{cc}$ via the two configurations of $(cc)[^3S_1]_{\bar{\textbf{3}}}$ and $(cc)[^1S_0]_{\textbf{6}}$, under the kinematic cuts listed in Equation (\ref{cuts}), are predicted to be: \\
for $E_e=60$ GeV,
\begin{eqnarray}
\sigma_{(cc)[^3S_1]_{\bar{\textbf{3}}}}&=&19.5^{+7.20}_{-5.20}~\textrm{pb}, \nonumber \\
\sigma_{(cc)[^1S_0]_{\textbf{6}}}&=&2.42^{+0.90}_{-0.65}~\textrm{pb},
\label{total60}
\end{eqnarray}
and for $E_e=140$ GeV,
\begin{eqnarray}
\sigma_{(cc)[^3S_1]_{\bar{\textbf{3}}}}&=&28.0^{+7.90}_{-6.60}~\textrm{pb}, \nonumber \\
\sigma_{(cc)[^1S_0]_{\textbf{6}}}&=&3.48^{+1.01}_{-0.82}~\textrm{pb},
\label{total140}
\end{eqnarray}
where the uncertainties are induced by varying $\mu_r(=\mu_f)$ from $0.5\mu_0$ to $2\mu_0$. From Equations (\ref{total60}) and (\ref{total140}), one can find:  halving or doubling $\mu_r$ and $\mu_f$ simultaneously around the default value of $\mu_0$ will arouse a $30\% \sim 40\%$ variation of the integrated cross section; the integrated cross sections corresponding to $E_e=140$ GeV is about $40\%$ larger in magnitude than that of $E_e=60$ GeV; the contribution via the configuration of $(cc)[^3S_1]_{\bar{\textbf{3}}}$ dominates over that of $(cc)[^1S_0]_{\textbf{6}}$, accounting for about $90\%$ in the total predictions.

Summing up the contributions of $(cc)[^3S_1]_{\bar{\textbf{3}}}$ and $(cc)[^1S_0]_{\textbf{6}}$, we can collect about $1.75(2.52) \times 10^{6}$ $\Xi_{cc}$ events corresponding to $E_e=60(140)$ GeV in one year (assuming $10^{7}$ seconds running time\footnote{Approximately, 1 year $\simeq \pi \times 10^{7}$ sceonds, but it is common that a collider only operates about $1/\pi$ year, i.e., $10^{34}\textrm{cm}^{-2}\textrm{s}^{-1} \simeq 10^{5}~\textrm{pb}^{-1}/\textrm{year}$.}), based on the up-to-date designed luminosity of LHeC $\mathcal{L}=0.8 \times 10^{34}\textrm{cm}^{-2}\textrm{s}^{-1}$. By further considering the decay chain of $\Xi_{cc}^{+} \to \Lambda_{c}^{+}K^{-}\pi^{+}$ ($\simeq 5\%$ \cite{Aaij:2013voa}) with the cascade decay $\Lambda_{c}^{+} \to pK^{-}\pi^{+}$($\simeq 5\%$ \cite{Aaij:2013voa}), or $\Xi_{cc}^{++} \to \Lambda_{c}^{+}K^{-}\pi^{+}\pi^{+}$ ($\simeq 10\%$ \cite{Aaij:2017ueg,Yu:2017zst}) with $\Lambda_{c}^{+} \to pK^{-}\pi^{+}$, we can accumulate about $1880$ reconstructed $\Xi^{+}_{cc}$ events, and $3760$ reconstructed $\Xi^{++}_{cc}$ events, corresponding to $E_e=60$ GeV. In the case of $E_e=140$ GeV, the numbers change to be about $2700$ and $5400$, respectively. Note that, throughout this paper the detection efficiency is assumed to be $100\%$. The thousands of reconstructed events per year strongly indicate the probability to hunt for the $\Xi^{+}_{cc}$ and $\Xi^{++}_{cc}$ events via the DIS at the LHeC.

For comparison, we now stop to test the prospect of observing the $\Xi_{cc}$ productions in DIS at the HERA. By adopting the same kinematic cuts as applied for the $J/\psi$ production \cite{DISexp2}, we have
\begin{eqnarray}
\sigma_{\Xi_{cc}}&=&3.61(3.69)~\textrm{pb},
\label{totalHERA}
\end{eqnarray}
corresponding to $E_e=$27.5 GeV and $E_P=820(920)$ GeV. According to the integrated luminosity of HERA, about $70~\textrm{pb}^{-1}$ in the years 1997-2000 \cite{DISexp2}, only less than 300 $\Xi_{cc}$ events can be generated there. Thus, it could be extremely hard to search for either $\Xi^{+}_{cc}$ or $\Xi^{++}_{cc}$ at the HERA by detecting their decaying into $\Lambda^{+}_c$ with subsequent $\Lambda^{+}_c \to pK^{-}\pi^{+}$.

\begin{table*}[htb]
\begin{center}
\caption{The integrated cross sections (unit: pb) of $\Xi_{cc}$ corresponding to different $Q^2$ bins (unit: $\textrm{GeV}^2$) under the kinematic cuts in Equation (\ref{cuts}). $N_{\Xi^{+}_{cc}}$ and $N_{\Xi^{++}_{cc}}$ denote the number of the $\Xi^{+}_{cc}$ and $\Xi^{++}_{cc}$ events per year, reconstructed by the decay chains of $\Xi_{cc}^{+} \to \Lambda_{c}^{+}K^{-}\pi^{+}$ with $\Lambda_{c}^{+} \to pK^{-}\pi^{+}$, and $\Xi_{cc}^{++} \to \Lambda_{c}^{+}K^{-}\pi^{+}\pi^{+}$ with $\Lambda_{c}^{+} \to pK^{-}\pi^{+}$.}
\begin{tabular}{lcccccccc}
\hline\hline
$~$ & $E_e=60~\textrm{GeV}$ & $~$ & $~$ & $E_e=140~\textrm{GeV}$ & $~$ & $~$ \\
\cline{2-4}\cline{5-7}
$Q^2_{\textrm{bin}}$ & $\sigma_{\Xi_{cc}}$ & $N_{\Xi^{+}_{cc}}$ & $N_{\Xi^{++}_{cc}}$ & $\sigma_{\Xi_{cc}}$ & $N_{\Xi^{+}_{cc}}$ & $N_{\Xi^{++}_{cc}}$ \\ \hline
$2-3.5$ & $6.14^{+2.07}_{-1.64}$ & $528^{+178}_{-141}$ & $1056^{+356}_{-282}$ & $8.67^{+2.16}_{-2.04}$ & $746^{+186}_{-175}$ & $1491^{+372}_{-351}$\\
$3.5-6.5$ & $5.73^{+2.06}_{-1.53}$ & $493^{+177}_{-132}$ & $~986^{+354}_{-263}$ & $8.16^{+2.21}_{-1.90}$ & $702^{+190}_{-163}$ & $1404^{+380}_{-327}$\\
$6.5-12$ & $4.32^{+1.67}_{-1.14}$ & $372^{+144}_{-~98}$ & $~743^{+287}_{-196}$ & $6.21^{+1.86}_{-1.46}$ & $534^{+160}_{-126}$ & $1068^{+320}_{-251}$\\
$12-20$ & $2.52^{+1.00}_{-0.67}$ & $217^{+~86}_{-~58}$ & $~433^{+172}_{-115}$ & $3.65^{+1.17}_{-0.88}$ & $314^{+101}_{-~76}$ & $~628^{+201}_{-151}$\\
$20-40$ & $2.05^{+0.82}_{-0.55}$ & $176^{+~71}_{-~47}$ & $~353^{+141}_{-~95}$ & $3.03^{+1.00}_{-0.74}$ & $261^{+~86}_{-~64}$ & $~527^{+172}_{-127}$\\
$40-100$ & $1.13^{+0.46}_{-0.31}$ & $~97^{+~40}_{-~27}$ & $~194^{+~79}_{-~53}$ & $1.72^{+0.59}_{-0.43}$ & $148^{+~51}_{-~37}$ & $~296^{+101}_{-~74}$\\
\hline \hline
\label{total cross section}
\end{tabular}
\end{center}
\end{table*}

\begin{figure}[!h]
\begin{center}
\hspace{0cm}\includegraphics[width=0.495\textwidth]{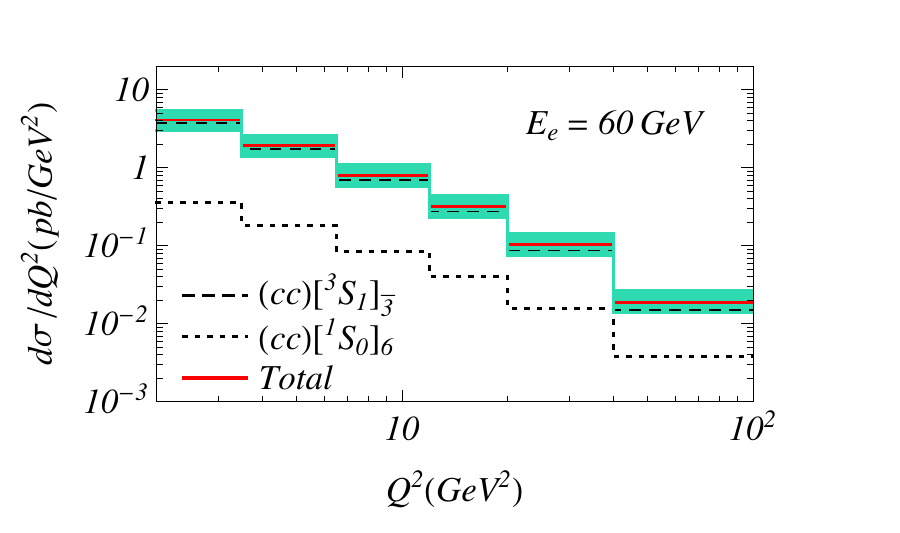}
\hspace{0cm}\includegraphics[width=0.495\textwidth]{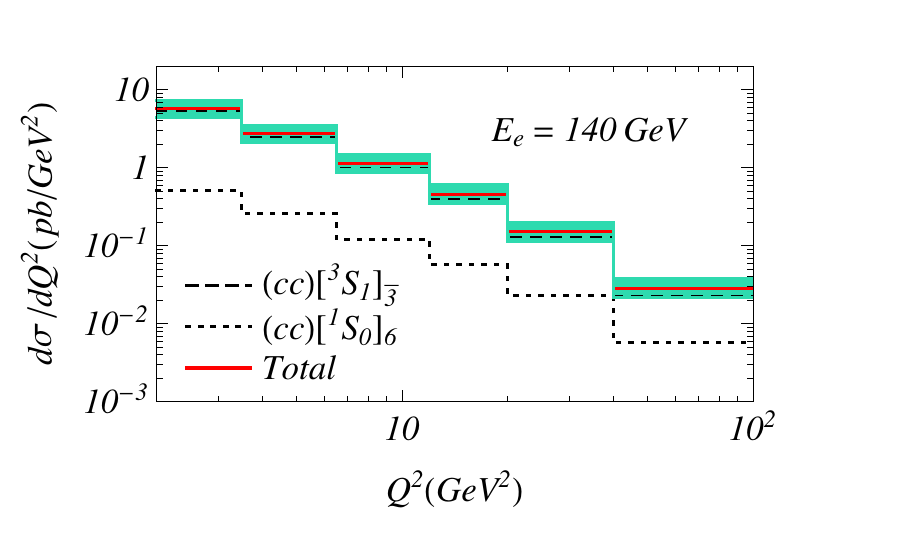}
\caption{\label{fig:Q2}
The differential cross sections as a function of $Q^2$ under the kinematic cuts in Equation (\ref{cuts}), corresponding to $E_e=60$ GeV and $E_e=140$ GeV, respectively. $``Total"$ denotes the sum of the contributions via $(cc)[^3S_1]_{\bar{\textbf{3}}}$ and $(cc)[^1S_0]_{\textbf{6}}$. The band is caused by the variation of $\mu_r(=\mu_f)$ from $0.5\mu_0$ to $2\mu_0$.}
\end{center}
\end{figure}

The virtuality of $\gamma^{*}$, representing by $Q^2$, is a key variable for the DIS process, thus it is interesting to investigate the dependence on it. For the integrated cross section, we choose some $Q^2$ bins between $2<Q^2<100~\textrm{GeV}^2$ to present their values in Table \ref{total cross section}. From the data in this table, we find that the dominant contributions are localized in the relatively small $Q^2$ scope, for instance, the proportion taken by the contribution of $2<Q^2<20~\textrm{GeV}^2$ to the total result ($2<Q^2<100~\textrm{GeV}^2$) is about $85\%$. Although the integrated cross sections of $20<Q^2<40~\textrm{GeV}^2$ and $40<Q^2<100~\textrm{GeV}^2$ are not as large as that of $2 \sim 20~\textrm{GeV}^2$, accounting for about $10\%$ and $5\%$, respectively, the corresponding hundreds of events also allow us to perform the measurements in those two $Q^2$ regions. Regarding the differential cross sections, we draw the $Q^2$ distributions in Figure \ref{fig:Q2}. Throughout this paper, we adopt the binning pattern, instead of continuous curves, to present the predictions on the differential cross sections, as the $J/\psi$ case at HERA \cite{DISexp2}. One can find that, as $Q^2$ increases, $d\sigma/dQ^2$ decreases rapidly; the contribution of the $(cc)[^1S_0]_{\textbf{6}}$ configuration becomes increasingly important.

\subsection{Differential cross sections}

\begin{figure}[!h]
\begin{center}
\hspace{0cm}\includegraphics[width=0.495\textwidth]{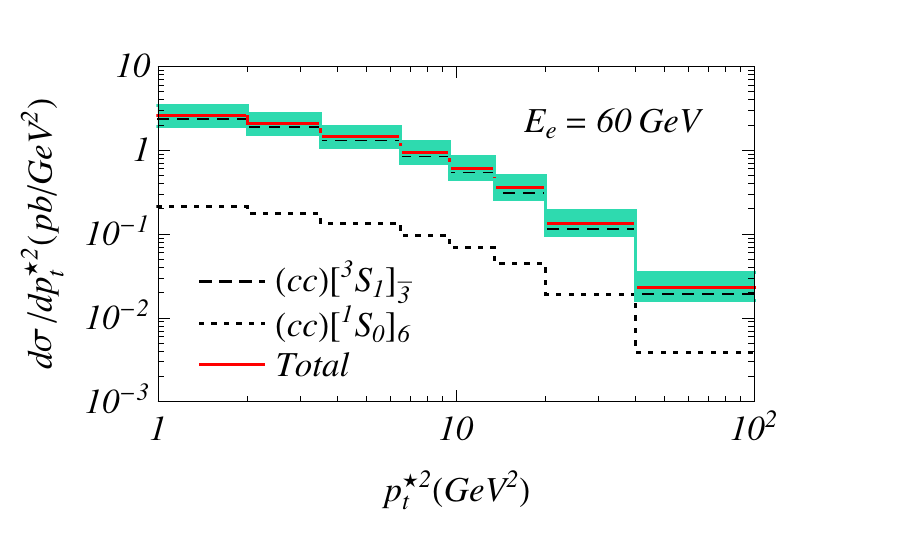}
\hspace{0cm}\includegraphics[width=0.495\textwidth]{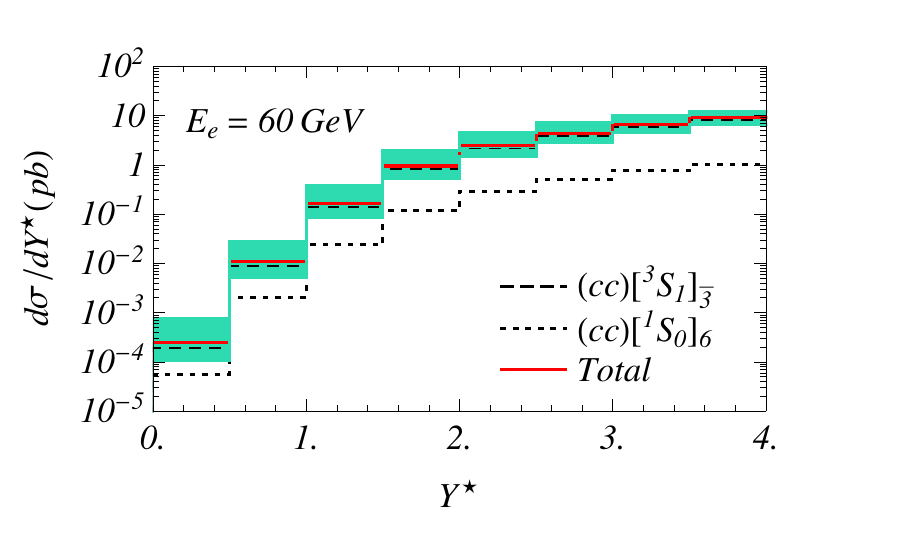}
\hspace{0cm}\includegraphics[width=0.495\textwidth]{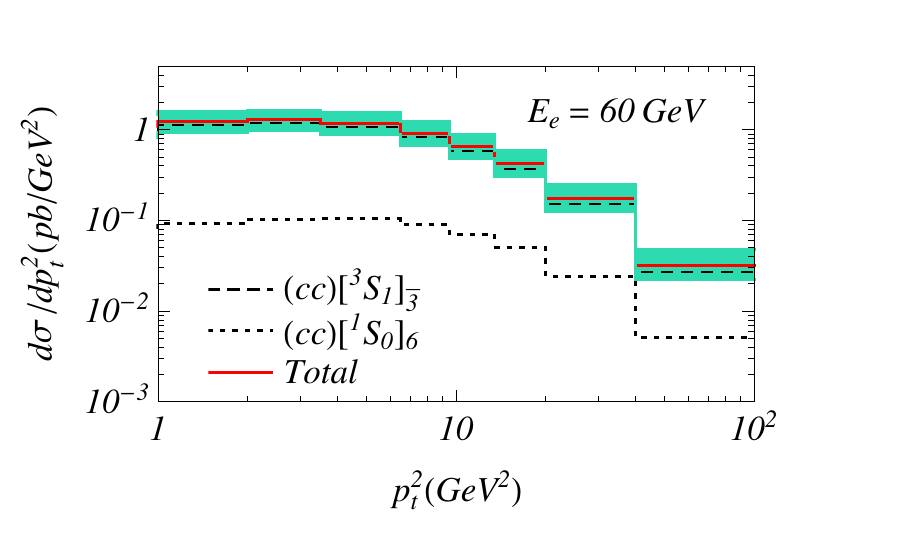}
\hspace{0cm}\includegraphics[width=0.495\textwidth]{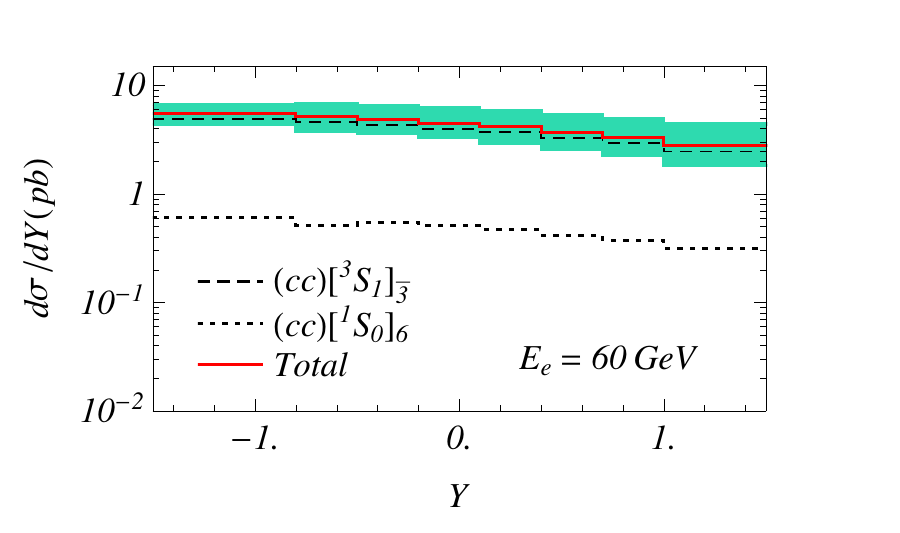}
\hspace{0cm}\includegraphics[width=0.495\textwidth]{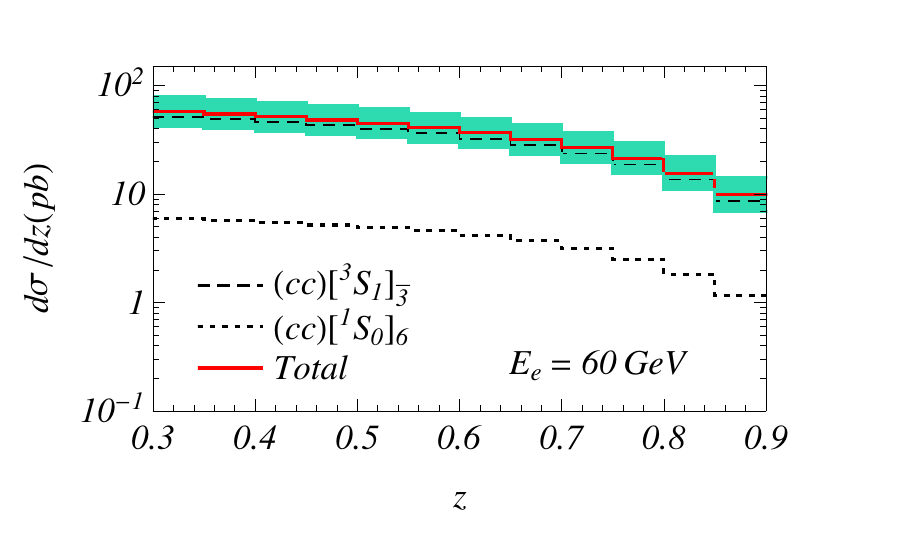}
\hspace{0cm}\includegraphics[width=0.495\textwidth]{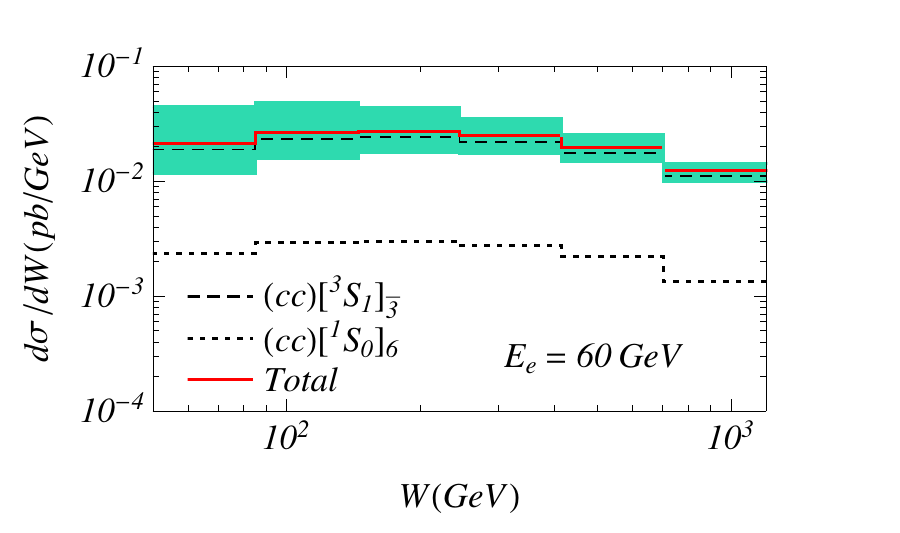}
\caption{\label{fig:dis60}
The distributions of $p_t^{2}(\Xi_{cc})$, $Y(\Xi_{cc})$, $p_t^{\star2}(\Xi_{cc})$, $Y^{\star}(\Xi_{cc})$, $W$, and $z(=\frac{p_{p} \cdot p_{\Xi_{cc}}}{p_{p} \cdot p_{\gamma^{*}}})$ under the kinematic cuts in Equation (\ref{cuts}), corresponding to $E_e=60$ GeV. $``Total"$ denotes the sum of the contributions via $(cc)[^3S_1]_{\bar{\textbf{3}}}$ and $(cc)[^1S_0]_{\textbf{6}}$. The band is caused by the variation of $\mu_r(=\mu_f)$ from $0.5\mu_0$ to $2\mu_0$.}
\end{center}
\end{figure}

\begin{figure}[!h]
\begin{center}
\hspace{0cm}\includegraphics[width=0.495\textwidth]{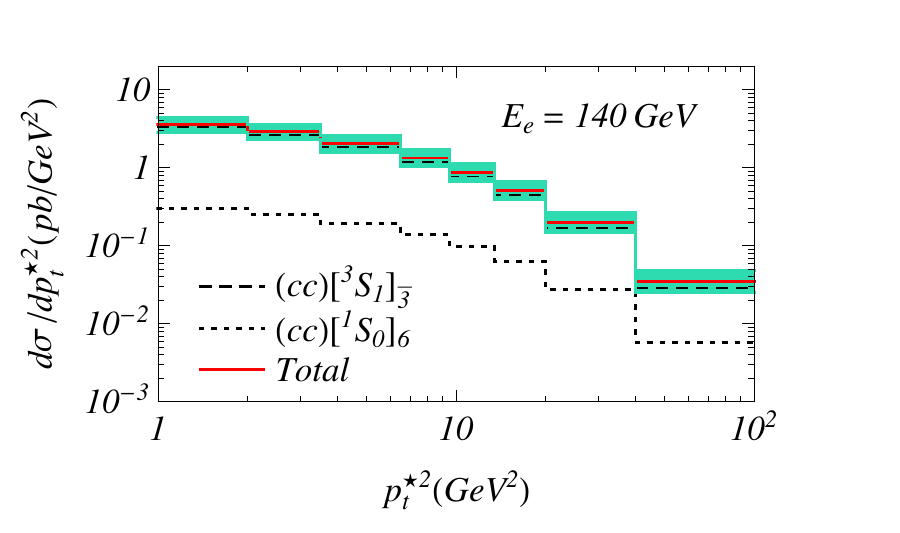}
\hspace{0cm}\includegraphics[width=0.495\textwidth]{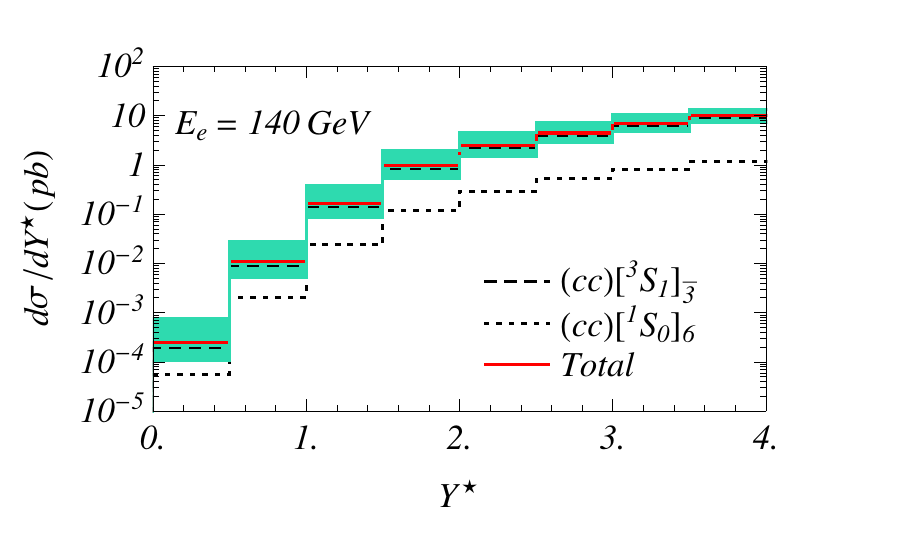}
\hspace{0cm}\includegraphics[width=0.495\textwidth]{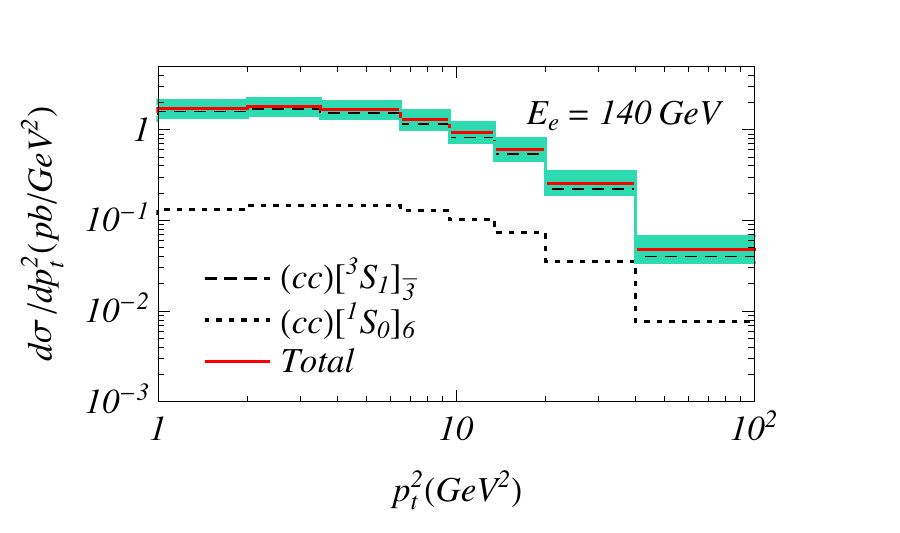}
\hspace{0cm}\includegraphics[width=0.495\textwidth]{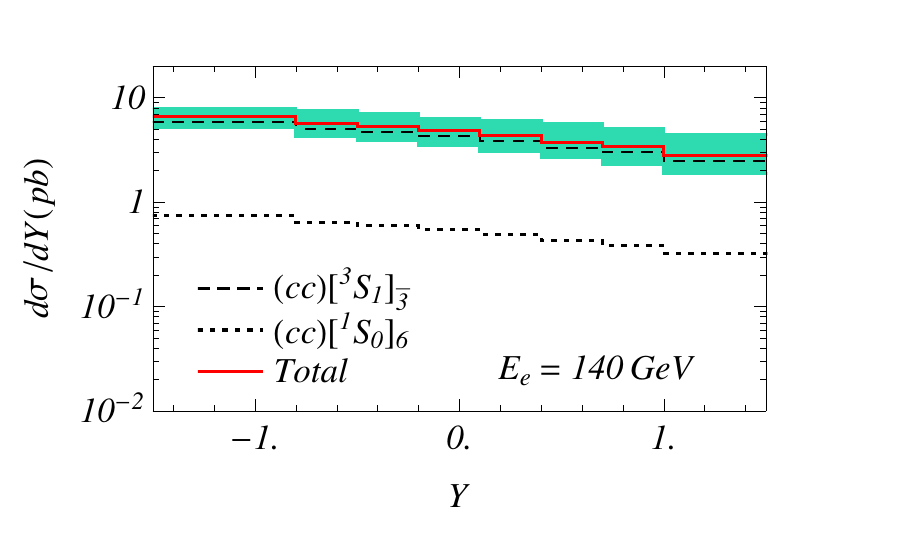}
\hspace{0cm}\includegraphics[width=0.495\textwidth]{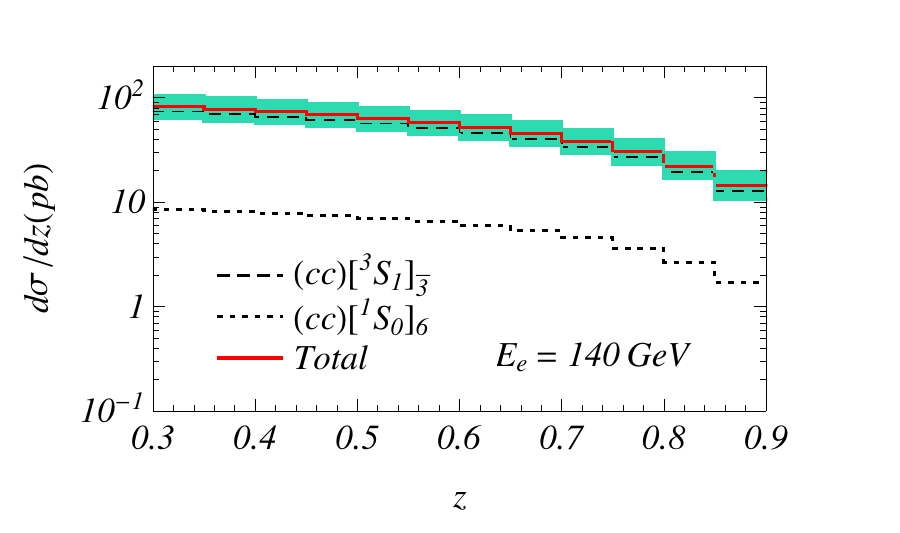}
\hspace{0cm}\includegraphics[width=0.495\textwidth]{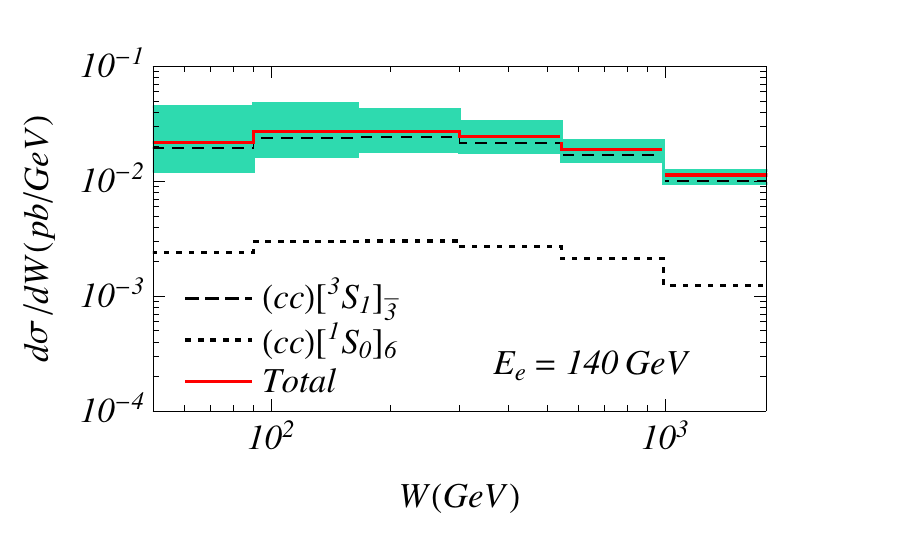}
\caption{\label{fig:dis140}
The distributions of $p_t^{2}(\Xi_{cc})$, $Y(\Xi_{cc})$, $p_t^{\star2}(\Xi_{cc})$, $Y^{\star}(\Xi_{cc})$, $W$, and $z(=\frac{p_{p} \cdot p_{\Xi_{cc}}}{p_{p} \cdot p_{\gamma^{*}}})$ under the kinematic cuts of Equation (\ref{cuts}), corresponding to $E_e=140$ GeV. $``Total"$ denotes the sum of the contributions via $(cc)[^3S_1]_{\bar{\textbf{3}}}$ and $(cc)[^1S_0]_{\textbf{6}}$. The band is caused by the variation of $\mu_r(=\mu_f)$ from $0.5\mu_0$ to $2\mu_0$.}
\end{center}
\end{figure}

As stated in section \ref{intro}, comparing to the hadro- and photoproduction, the production in DIS allow us to measure more varieties of physical observables. Therefore, to serve as a preliminary, useful reference awaiting for the future measurements by LHeC, in addition to $d\sigma/dQ^2$, we further provide the differential cross sections of $\Xi_{cc}$ with respect to $p_{t}^{2}$, $Y$, $p_{t}^{\star2}$, and $Y^{\star}$, as well as the $W$, $z$ distributions, in Figures \ref{fig:dis60} and \ref{fig:dis140}, which correspond to $E_e=60$ GeV and $E_e=140$ GeV, respectively. Following the conventions of HERA, the forward direction of $Y^{\star}$ is defined as that of the incident virtual photo; $Y$ is taken to be positive in the direction of the incoming proton. As can be seen in the two figures,
\begin{itemize}
\item[1)]
As $p_t^{\star2}$ goes up, the differential cross sections continuously decrease, and the relative significance of the $(cc)[^1S_0]_{\textbf{6}}$ configuration increases gradually. With regard to the $p_t^2$ distribution, the situation is similar to the $p_t^{\star2}$ case, except that there is a small peak around $p_t^2 \sim 2.5~\textrm{GeV}^2$.
\item[2)]
For $Y^{\star}$ (rapidity) distribution, the differential cross section is seriously asymmetric. To be specific, the value of $d\sigma/dY^{\star}$ at $Y^{\star}=4$ is about five orders of magnitudes larger than that at $Y^{\star}=0$, which indicates, in the $\gamma^{*}p$ CM frame, $\Xi_{cc}$ is much more likely to be generated in the direction of the virtual photon rather than that of the incoming proton. The $Y$ distribution is also asymmetric, for example, $\frac{d\sigma}{dY}|_{Y=-1.5}$ is about two times larger in magnitude than $\frac{d\sigma}{dY}|_{Y=1.5}$, from which we can learn that, in the laboratory frame, more $\Xi_{cc}$ events will be generated along the direction of the incident electron.
\item[3)]
For $W$ distribution, the available $W$ values are scatted from tens to thousands, and the peak of the differential cross section is localized at the $W$ range of $100 \sim 150$ GeV. Regarding the $z$ distribution, $d\sigma/dz$ corresponding to small $z$ is several times bigger than that for large $z$. This inequality implies that the small (mid) $z$ region is the main source of the $\Xi_{cc}$ events.
\end{itemize}

In addition to LHeC, the forthcoming electron-ion collider (EIC) \cite{EIC} that is planned to be built at Brookhaven Laboratory (USA) is also helpful to study the DIS processes. Thus, at last we carry out an estimation on the $\Xi_{cc}$ production rates under the EIC running energy and luminosity. Taking the same kinematic cuts in Equation (\ref{cuts}), the integrated cross section of $\Xi_{cc}$ in deeply inelastic $ep$ scattering at EIC is predicted to be  
\begin{eqnarray}
\sigma_{\Xi_{cc}}&=&0.48(1.39)~\textrm{pb},
\end{eqnarray}
corresponding to $E_e=$21 GeV and $E_P=100(250)$ GeV. Taking a conservative choice of the designed luminosity of EIC, $\mathcal{L} \sim 10^{34}\textrm{cm}^{-2}\textrm{s}^{-1}$, about $50(150)$ reconstructed $\Xi^{+}_{cc}$ events and $100(300)$ reconstructed $\Xi^{++}_{cc}$ events can be accumulated in one operation year (assuming $10^7$ seconds), which indicates the EIC could also be a helpful laboratory to study the $\Xi_{cc}$ related DIS processes.
\section{Summary}\label{sum}

In this manuscript, we have carried out the studies on the productions of the doubly charmed baryons, $\Xi^{+}_{cc}$ and $\Xi^{++}_{cc}$, in deeply inelastic $ep$ scattering at the LHeC with $E_e=60,140$ GeV and $E_p=7000$ GeV. Imposing the kinematic cuts as adopted in observing the inclusive $J/\psi$ productions in DIS at the HERA, about 1880 (2700) $\Xi^{+}_{cc}$ events and 3760(5400) $\Xi^{++}_{cc}$ events can be accumulated in one year by hunting for the decay chains of $\Xi_{cc}^{+} \to \Lambda_{c}^{+}K^{-}\pi^{+}$ with the cascade decay $\Lambda_{c}^{+} \to pK^{-}\pi^{+}$, and $\Xi_{cc}^{++} \to \Lambda_{c}^{+}K^{-}\pi^{+}\pi^{+}$ with $\Lambda_{c}^{+} \to pK^{-}\pi^{+}$. We also provide the predictions on the distributions of $Q^2$, $p_t^{2}$, $Y$, $p_t^{\star2}$, $Y^{\star}$, $W$, and $z$, which may serve as a useful reference for the future measurements at LHeC. Thousands of reconstructed events manifest the possibility of observing $\Xi^{+}_{cc}$ and $\Xi^{++}_{cc}$ baryons via the DIS production process at the LHeC. Thus we think, in addition to LHC, the LHeC can also be a good laboratory for investigating the properties of the doubly heavy baryons.

\acknowledgments
We thank Dr. Hong-Fei Zhang for helpful discussions. This work is supported in part by the Natural Science Foundation of China under the Grant No. 11647113,  No. 11705034 and No. 11625520, by the Project for Young Talents Growth of Guizhou Provincial Department of Education under Grant No.KY[2017]135, and by the Project of GuiZhou Provincial Department of Science and Technology under Grant No. QKHJC[2020]1Y035. \\

\providecommand{\href}[2]{#2}\begingroup\raggedright


\begin{thebibliography}{10}

\bibitem{GellMann:1964nj}
  M.~Gell-Mann, {\it {A Schematic Model of Baryons and Mesons}},  {\em Phys.\ Lett.} {\bf 8} (1964) 214.

\bibitem{Zweig}
  G.~Zweig, {\it {An SU(3) model for strong interaction symmetry and its breaking}}, CERN-TH-401.

\bibitem{Ebert:1996ec}
  D.~Ebert, R.~N.~Faustov, V.~O.~Galkin, A.~P.~Martynenko and V.~A.~Saleev,
  {\it {Heavy baryons in the relativistic quark model}},
  {\em Z.\ Phys.}{\bf \ C76} (1997) 111, [\href{https://arxiv.org/abs/hep-ph/9607314}{{\tt hep-ph/9607314}}].

\bibitem{Gerasyuta:1999pc}
  S.~M.~Gerasyuta and D.~V.~Ivanov,
  {\it {Charmed baryons in bootstrap quark model}},
  {\em Nuovo Cim.} {\bf A112} (1999) 261, [\href{https://arxiv.org/abs/hep-ph/0101310}{{\tt hep-ph/0101310}}].

\bibitem{Itoh:2000um}
  C.~Itoh, T.~Minamikawa, K.~Miura and T.~Watanabe,
  {\it {Doubly charmed baryon masses and quark wave functions in baryons}},
  {\em Phys.\ Rev.} {\bf D61} (2000) 057502.

\bibitem{Mattson:2002vu}
  {\bf SELEX} Collaboration, M.~Mattson {\it et al.},
  {\it {First Observation of the Doubly Charmed Baryon $\Xi^+_{cc}$}},
  {\em Phys.\ Rev.\ Lett.} {\bf 89} (2002) 112001, [\href{https://arxiv.org/abs/hep-ex/0208014}{{\tt hep-ex/0208014}}].

\bibitem{Ocherashvili:2004hi}
  {\bf SELEX} Collaboration, A.~Ocherashvili {\it et al.},
  {\it {Confirmation of the double charm baryon Xi+(cc)(3520) via its decay to p D+ K-}},
 {\em Phys.\ Lett.} {\bf B628} (2005) 18, [\href{https://arxiv.org/abs/hep-ex/0406033}{{\tt hep-ex/0406033}}].

\bibitem{Aubert:2006qw}
  {\bf BaBar} Collaboration, B.~Aubert {\it et al.},
  {\it {Search for doubly charmed baryons Xi(cc)+ and Xi(cc)++ in BABAR}},
  {\em Phys.\ Rev.} {\bf D74} (2006) 011103, [\href{https://arxiv.org/abs/hep-ex/0605075}{{\tt hep-ex/0605075}}].

\bibitem{Chistov:2006zj}
  {\bf Belle} Collaboration, R.~Chistov {\it et al.},
  {\it {Observation of new states decaying into Lambda(c)+ K- pi+ and Lambda(c)+ K0(S) pi-}},
  {\em Phys.\ Rev.\ Lett.} {\bf 97} (2006) 162001, [\href{https://arxiv.org/abs/hep-ex/0606051}{{\tt hep-ex/0606051}}].

\bibitem{Kato:2013ynr}
  {\bf Belle} Collaboration, Y.~Kato {\it et al.},
  {\it {Search for doubly charmed baryons and study of charmed strange baryons at Belle}},
  {\em Phys.\ Rev.} {\bf D89} (2014),  052003, [\href{https://arxiv.org/abs/1312.1026}{{\tt arXiv:1312.1026}}].

\bibitem{Ratti:2003ez}
  S.~P.~Ratti,
  {\it {New results on c-baryons and a search for cc-baryons in FOCUS}},
  {\em Nucl.\ Phys.\ Proc.\ Suppl.} {\bf 115} (2003) 33.

\bibitem{Aaij:2013voa}
  {\bf LHCb} Collaboration, R.~Aaij {\it et al.},
  {\it {Search for the doubly charmed baryon $\Xi_{cc}^+$}},
  {\em JHEP} {\bf 1312} (2013) 090, [\href{https://arxiv.org/abs/1310.2538}{{\tt arXiv:1310.2538}}].

\bibitem{Aaij:2017ueg}
  {\bf LHCb} Collaboration, R.~Aaij {\it et al.}],
  {\it {Observation of the doubly charmed baryon $\Xi_{cc}^{++}$}},
  {\em Phys.\ Rev.\ Lett.} {\bf 119} (2017) , 112001, [\href{https://arxiv.org/abs/1707.01621}{{\tt arXiv:1707.01621}}].

\bibitem{Aaij:2018gfl}
  {\bf LHCb} Collaboration, R.~Aaij {\it et al.},
  {\it {First Observation of the Doubly Charmed Baryon Decay $\Xi_{cc}^{++}\rightarrow \Xi_{c}^{+}\pi^{+}$}},
  {\em Phys.\ Rev.\ Lett.} {\bf 121} (2018) , 162002, [\href{https://arxiv.org/abs/1807.01919}{{\tt arXiv:1807.01919}}].

\bibitem{Aaij:2018wzf}
  {\bf LHCb} Collaboration, R.~Aaij {\it et al.},
  {\it {Measurement of the Lifetime of the Doubly Charmed Baryon $\Xi_{cc}^{++}$}},
  {\em Phys.\ Rev.\ Lett.} {\bf 121} (2018) , 052002, [\href{https://arxiv.org/abs/1806.02744}{{\tt arXiv:1806.02744}}].

\bibitem{Aaij:2019jfq}
  {\bf LHCb} Collaboration, R.~Aaij {\it et al.},
  {\it {Search for the doubly charmed baryon $\Xi_{cc}^+$}},
  {\em Sci.\ China Phys.\ Mech.\ Astron.} {\bf 63} (2020), 221062, [\href{https://arxiv.org/abs/1909.12273}{{\tt arXiv:1909.12273}}].
  
\bibitem{Aaij:2019zxa}
  {\bf LHCb} Collaboration, R.~Aaij {\it et al.},
  {\it {Measurement of $\mathit{\Xi}_{cc}^{++}$ production in $pp$ collisions at $\sqrt{s}=13$ TeV}},
  {\em Chin.\ Phys.} {\bf C44} (2020) , 022001, [\href{https://arxiv.org/abs/1910.11316}{{\tt arXiv:1910.11316}}].

\bibitem{literature21}
  S.~J.~Brodsky, S.~Groote and S.~Koshkarev,
  {\it {Resolving the SELEX-LHCb double-charm baryon conflict: the impact of intrinsic heavy-quark hadroproduction and supersymmetric light-front holographic QCD}},
  {\em Eur.\ Phys.\ J.} {\bf C78} (2018), 483, [\href{https://arxiv.org/abs/1709.09903}{{\tt arXiv:1709.09903}}].

\bibitem{literature24}
  C.~H.~Chang, J.~P.~Ma, C.~F.~Qiao and X.~G.~Wu,
  {\it {Hadronic production of the doubly charmed baryon Xi(cc) with intrinsic charm}},
  {\em J.\ Phys.\ G} {\bf 34} (2007) 845, [\href{https://arxiv.org/abs/hep-ph/0610205}{{\tt hep-ex/0610205}}].
  
\bibitem{literature1}
  A.~F.~Falk, M.~E.~Luke, M.~J.~Savage and M.~B.~Wise,
  {\it {Heavy quark fragmentation to baryons containing two heavy quarks}},
  {\em Phys.\ Rev.} {\bf D49} (1994) 555, [\href{https://arxiv.org/abs/hep-ph/9305315}{{\tt hep-ex/9305315}}].

\bibitem{literature2}
  V.~V.~Kiselev, A.~K.~Likhoded and M.~V.~Shevlyagin,
  {\it {Double charmed baryon production at B factory}},
  {\em Phys.\ Lett.} {\bf B332} (1994) 411, [\href{https://arxiv.org/abs/hep-ph/9408407}{{\tt hep-ex/9408407}}].

\bibitem{literature3}
  S.~P.~Baranov,
  {\it {On the production of doubly flavored baryons in p p, e p and gamma gamma collisions}},
  {\em Phys.\ Rev.} {\bf D54} (1996) 3228.

\bibitem{literature4}
  A.~V.~Berezhnoy, V.~V.~Kiselev, A.~K.~Likhoded and A.~I.~Onishchenko,
  {\it {Doubly charmed baryon production in hadronic experiments}},
  {\em Phys.\ Rev.} {\bf D57} (1998) 4385, [\href{https://arxiv.org/abs/hep-ph/9710339}{{\tt hep-ex/9710339}}].

\bibitem{literature5}
  D.~A.~Gunter and V.~A.~Saleev,
  {\it {Hadronic production of doubly charmed baryons via charm excitation in proton}},
  {\em Phys.\ Rev.} {\bf D64} (2001) 034006, [\href{https://arxiv.org/abs/hep-ph/0104173}{{\tt hep-ex/0104173}}].

\bibitem{literature51}
  V.~V.~Kiselev, A.~K.~Likhoded and M.~V.~Shevlyagin,
  {\it {Production of doubly charmed baryons at energy s**(1/2) = 10.58-GeV}},
  {\em Phys.\ Atom.\ Nucl.} {\bf 58} (1995) 1018, [{\em Yad.\ Fiz.}  {\bf 58} (1995) 1092].

\bibitem{literature52}
  A.~V.~Berezhnoy and A.~K.~Likhoded,
  {\it {Quark-hadron duality and production of charmonia and doubly charmed baryons in e+ e- annihilation}},
  {\em Phys.\ Atom.\ Nucl.}  {\bf 70} (2007) 478, [\href{https://arxiv.org/abs/hep-ph/0602041}{{\tt hep-ex/0602041}}].

\bibitem{literature6}
  V.~V.~Braguta, V.~V.~Kiselev and A.~E.~Chalov,
  {\it {Pair production of doubly heavy diquarks}},
  {\em Phys.\ Atom.\ Nucl.} {\bf 65} (2002) 1537 [{\em Yad.\ Fiz.} {\bf 65} (2002) 1575].

\bibitem{literature7}
  J.~P.~Ma and Z.~G.~Si,
  {\it {Factorization approach for inclusive production of doubly heavy baryon}},
  {\em Phys.\ Lett.} {\bf B568} (2003) 135, [\href{https://arxiv.org/abs/hep-ph/0305079}{{\tt hep-ex/0305079}}].

\bibitem{literature8}
  E.~Braaten, M.~Kusunoki, Y.~Jia and T.~Mehen,
  {\it {Lambda+(c) / Lambda-(c) asymmetry in hadroproduction from heavy quark recombination}},
  {\em Phys.\ Rev.} {\bf D70} (2004) 054021, [\href{https://arxiv.org/abs/hep-ph/0304280}{{\tt hep-ex/0304280}}].

\bibitem{literature9}
  S.~Y.~Li, Z.~G.~Si and Z.~J.~Yang,
  {\it {Doubly heavy baryon production at gamma gamma collider}},
  {\em Phys.\ Lett.} {\bf B648} (2007) 284, [\href{https://arxiv.org/abs/hep-ph/0701212}{{\tt hep-ex/0701212}}].

\bibitem{literature10}
  Z.~J.~Yang and T.~Yao,
  {\it {Doubly heavy baryon production at polarized photon collider}},
  {\em Chin.\ Phys.\ Lett.} {\bf 24} (2007) 3378, [\href{https://arxiv.org/abs/0710.0051}{{\tt arXiv:0710.0051}}].

\bibitem{literature11}
  H.~Y.~Bi, R.~Y.~Zhang, X.~G.~Wu, W.~G.~Ma, X.~Z.~Li and S.~Owusu,
  {\it {Photoproduction of doubly heavy baryon at the LHeC}},
  {\em Phys.\ Rev.} {\bf D95} (2017), 074020, [\href{https://arxiv.org/abs/1702.07181}{{\tt arXiv:1702.07181}}].

\bibitem{literature12}
  J.~W.~Zhang, X.~G.~Wu, T.~Zhong, Y.~Yu and Z.~Y.~Fang,
  {\it {Hadronic Production of the Doubly Heavy Baryon $\Xi_{bc}$ at LHC}},
  {\em Phys.\ Rev.} {\bf D83} (2011) 034026, [\href{https://arxiv.org/abs/1101.1130}{{\tt arXiv:1101.1130}}].

\bibitem{literature13}
  J.~Jiang, X.~G.~Wu, Q.~L.~Liao, X.~C.~Zheng and Z.~Y.~Fang,
  {\it {Doubly Heavy Baryon Production at A High Luminosity $e^+ e^-$ Collider}},
  {\em Phys.\ Rev.} {\bf D86} (2012) 054021, [\href{https://arxiv.org/abs/1208.3051}{{\tt arXiv:1208.3051}}].

\bibitem{literature14}
  J.~Jiang, X.~G.~Wu, S.~M.~Wang, J.~W.~Zhang and Z.~Y.~Fang,
  {\it {A Further Study on the Doubly Heavy Baryon Production around the $Z^0$ Peak at A High Luminosity $e^+ e^-$ Collider}},
  {\em Phys.\ Rev.} {\bf D87} (2013), 054027, [\href{https://arxiv.org/abs/1302.0601}{{\tt arXiv:1302.0601}}].

\bibitem{literature15}
  A.~P.~Martynenko and A.~M.~Trunin,
  {\it {Relativistic corrections to the pair double heavy diquark production in $e^+ e^-$ annihilation}},
  {\em Phys.\ Rev.} {\bf D89} (2014), 014004, [\href{https://arxiv.org/abs/1308.3998}{{\tt arXiv:1308.3998}}].

\bibitem{literature16}
  G.~Chen, X.~G.~Wu, Z.~Sun, Y.~Ma and H.~B.~Fu,
 {\it {Photoproduction of doubly heavy baryon at the ILC}},
 {\em JHEP} {\bf 1412} (2014) 018, [\href{https://arxiv.org/abs/1408.4615}{{\tt arXiv:1408.4615}}].

\bibitem{literature17}
  Z.~J.~Yang and X.~X.~Zhao,
  {\it {The Production of $\Xi_{bb}$ at Photon Collider}},
  {\em Chin.\ Phys.\ Lett.} {\bf 31} (2014), 091301, [\href{https://arxiv.org/abs/1408.5584}{{\tt arXiv:1408.5584}}].

\bibitem{literature18}
  Z.~J.~Yang, P.~F.~Zhang and Y.~J.~Zheng,
  {\it {Doubly Heavy Baryon Production in $e^{+}e^{-}$ Annihilation}},
  {\em Chin.\ Phys.\ Lett.} {\bf 31} (2014) 051301.

\bibitem{literature19}
  A.~P.~Martynenko and A.~M.~Trunin,
  {\it {Pair double heavy diquark production in high energy proton–proton collisions}},
  {\em Eur.\ Phys.\ J.} {\bf C75} (2015), 138, [\href{https://arxiv.org/abs/1405.0969}{{\tt arXiv:1405.0969}}].

\bibitem{literature20}
  W.~K.~Lai and A.~K.~Leibovich,
  {\it {$\Lambda{_c}^{+}/\Lambda_{c}^{-}$ and $\Lambda_{b}^{0}/\bar{\Lambda}_{b}^0$ production asymmetry at the LHC from heavy quark recombination}},
  {\em Phys.\ Rev.} {\bf D91} (2015), 054022, [\href{https://arxiv.org/abs/1410.2091}{{\tt arXiv:1410.2091}}].

\bibitem{literature201}
  S.~Koshkarev and V.~Anikeev,
  {\it {Production of the doubly charmed baryons at the SELEX experiment - The double intrinsic charm approach}},
  {\em Phys.\ Lett.} {\bf B765} (2017) 171, [\href{https://arxiv.org/abs/1605.03070}{{\tt arXiv:1605.03070}}].

\bibitem{literature202}
  S.~Koshkarev,
  {\it {Production of the Doubly Heavy Baryons, $B_c$ Meson and the All-charm Tetraquark at AFTER@LHC with Double Intrinsic Heavy Mechanism}},''
  {\em Acta Phys.\ Polon.} {\bf B48} (2017) 163, [\href{https://arxiv.org/abs/1610.06125}{{\tt arXiv:1610.06125}}].

\bibitem{literature203}
  S.~Groote and S.~Koshkarev,
  {\it {Production of doubly charmed baryons nearly at rest}},
  {\em Eur.\ Phys.\ J.} {\bf C77} (2017), 509, [\href{https://arxiv.org/abs/1704.02850}{{\tt arXiv:1704.02850}}].

\bibitem{literature22}
  X.~Yao and B.~Müller,
  {\it {Doubly charmed baryon production in heavy ion collisions}},
  {\em Phys.\ Rev.} {\bf D97} (2018), 074003, [\href{https://arxiv.org/abs/1801.02652}{{\tt arXiv:1801.02652}}].

\bibitem{literature23}
  C.~H.~Chang, C.~F.~Qiao, J.~X.~Wang and X.~G.~Wu,
  {\it {Estimate of the hadronic production of the doubly charmed baryon Xi(cc) under GM-VFN scheme}},
  {\em Phys.\ Rev.} {\bf D73} (2006) 094022, [\href{https://arxiv.org/abs/hep-ph/0601032}{{\tt hep-ex/0601032}}].

\bibitem{literature25}
  G.~Chen, X.~G.~Wu, J.~W.~Zhang, H.~Y.~Han and H.~B.~Fu,
  {\it {Hadronic production of $\Xi_{cc}$ at a fixed-target experiment at the LHC}},
  {\em Phys.\ Rev.} {\bf D89} (2014), 074020, [\href{https://arxiv.org/abs/1401.6269}{{\tt arXiv:1401.6269}}].

\bibitem{literature26}
  X.~C.~Zheng, C.~H.~Chang and Z.~Pan,
  {\it {Production of doubly heavy-flavored hadrons at $e^+e^-$ colliders}},
  {\em Phys.\ Rev.} {\bf D93} (2016), 034019, [\href{https://arxiv.org/abs/1510.06808}{{\tt arXiv:1510.06808}}].

\bibitem{literature27}
  G.~Chen, C.~H.~Chang and X.~G.~Wu,
  {\it {Hadronic production of the doubly charmed baryon via the proton–nucleus and the nucleus–nucleus collisions at the RHIC and LHC}},
  {\em Eur.\ Phys.\ J.} {\bf C78} (2018), 801, [\href{https://arxiv.org/abs/1808.03174}{{\tt arXiv:1808.03174}}].

\bibitem{literature271}
  A.~V.~Berezhnoy, I.~N.~Belov and A.~K.~Likhoded,
  {\it {Production of doubly charmed baryons with the excited heavy diquark at LHC}},
  {Int.\ J.\ Mod.\ Phys.} {\bf A34} (2019), 1950038, [\href{https://arxiv.org/abs/1811.07382}{{\tt arXiv:1811.07382}}].

\bibitem{literature28}
  G.~Chen, X.~G.~Wu and S.~Xu,
  {\it {Impacts of the intrinsic charm content of the proton on the $\Xi_{cc}$ hadroproduction at a fixed target experiment at the LHC}},''
  {\em Phys.\ Rev.} {\bf D100} (2019), 054022, [\href{https://arxiv.org/abs/1903.00722}{{\tt arXiv:1903.00722}}].

\bibitem{literature29}
  X.~G.~Wu,
  {\it {A new search for the doubly charmed baryon $\Xi_{cc}^+$ at the LHC}},
  {\em Sci.\ China Phys.\ Mech.\ Astron.}  {\bf 63} (2020) 221063, [\href{https://arxiv.org/abs/1912.01953}{{\tt arXiv:1912.01953}}].

\bibitem{literature30}
  J.~J.~Niu, L.~Guo, H.~H.~Ma, X.~G.~Wu and X.~C.~Zheng,
  {\it {Production of semi-inclusive doubly heavy baryons via top-quark decays}},''
  {\em Phys.\ Rev.} {\bf D98} (2018), 094021, [\href{https://arxiv.org/abs/1810.03834}{{\tt arXiv:1810.03834}}].

\bibitem{literature31}
  J.~J.~Niu, L.~Guo, H.~H.~Ma and X.~G.~Wu,
  {\it {Production of doubly heavy baryons via Higgs boson decays}},
  {\em Eur.\ Phys.\ J.} {\bf C79} (2019), 339, [\href{https://arxiv.org/abs/1904.02339}{{\tt arXiv:1904.02339}}].

\bibitem{Chang:2007pp}
  C.~H.~Chang, J.~X.~Wang and X.~G.~Wu,
  {\it GENXICC: A Generator for hadronic production of the double heavy baryons Xi(cc), Xi(bc) and Xi(bb)},
  {\em Comput.\ Phys.\ Commun.}\  {\bf 177} (2007), 467, [\href{https://arxiv.org/abs/hep-ph/0702054}{{\tt arXiv:0702054}}].

\bibitem{Chang:2009va}
  C.~H.~Chang, J.~X.~Wang and X.~G.~Wu,
  {\it GENXICC2.0: An Upgraded Version of the Generator for Hadronic Production of Double Heavy Baryons Xi(cc), Xi(bc) and Xi(bb)},
  {\em Comput.\ Phys.\ Commun.}\  {\bf 181} (2010), 1144, [\href{https://arxiv.org/abs/0910.4462}{{\tt arXiv:0910.4462}}].

\bibitem{Wang:2012vj}
  X.~Y.~Wang and X.~G.~Wu,
  {\it GENXICC2.1: An Improved Version of GENXICC for Hadronic Production of Doubly Heavy Baryons},
  {\em Comput.\ Phys.\ Commun.}\  {\bf 184} (2013), 1070, [\href{https://arxiv.org/abs/1210.3458}{{\tt arXiv:1210.3458}}].

\bibitem{Jpsi1}
  R.~Baier and R.~Ruckl,
  {\it {On Inelastic Leptoproduction of Heavy Quarkonium States}},
  {\em Nucl.\ Phys.} {\bf B201} (1982) 1.

\bibitem{Jpsi2}
  J.~G.~Korner, J.~Cleymans, M.~Kuroda and G.~J.~Gounaris,
  {\it {Azimuthal Dependence of Deep Inelastic Heavy Resonance Production}},
  {\em Phys.\ Lett.}  {\bf B114} (1982) 195.

\bibitem{Jpsi3}
  J.~P.~Guillet,
  {\it {A Way to Measure the Spin Dependent Distribution of the Gluon}},
  {\em Z.\ Phys.} {\bf C39} (1988) 75.

\bibitem{Jpsi4}
  H.~Merabet, J.~F.~Mathiot and R.~Mendez-Galain,
  {\it {Inelastic leptoproduction of J / psi and the gluon distribution in the nucleon}},
  {\em Z.\ Phys.} {\bf C62} (1994) 639.

\bibitem{Jpsi5}
  S.~Fleming and T.~Mehen,
  {\it {Leptoproduction of $J/\psi$}},
  {\em Phys.\ Rev.} {\bf D57} (1998) 1846, [\href{https://arxiv.org/abs/hep-ph/9707365}{{\tt hep-ex/9707365}}].

\bibitem{Jpsi6}
  F.~Yuan and K.~T.~Chao,
  {\it {Polarized $J/\psi$ production in deep inelastic scattering at HERA}},
  {\em Phys.\ Rev.} {\bf D63} (2001) 034017, Erratum: [{\em Phys.\ Rev.} {\bf D66} (2002) 079902], [\href{https://arxiv.org/abs/hep-ph/0008301}{{\tt hep-ex/0008301}}].

\bibitem{Jpsi7}
  B.~A.~Kniehl and L.~Zwirner,
  {\it {$J/\psi$ inclusive production in $e p$ deep inelastic scattering at DESY HERA}},
  {\em Nucl.\ Phys.} {\bf B621} (2002) 337, [\href{https://arxiv.org/abs/hep-ph/0112199}{{\tt hep-ex/0112199}}].

\bibitem{Jpsi8}
  H.~F.~Zhang and Z.~Sun,
  {\it {Leptonic current structure and azimuthal asymmetry in deeply inelastic scattering}},
  {\em Phys.\ Rev.} {\bf D96} (2017), 034002, [\href{https://arxiv.org/abs/1701.08728}{{\tt arXiv:1701.08728}}].

\bibitem{Jpsi9}
  Z.~Sun and H.~F.~Zhang,
  {\it {QCD leading order study of the $J/\psi$ leptoproduction at HERA within the nonrelativistic QCD framework}},''
  {\em Eur.\ Phys.\ J.} {\bf C77} (2017), 744, [\href{https://arxiv.org/abs/1702.02097}{{\tt arXiv:1702.02097}}].

\bibitem{Jpsi10}
  Z.~Sun and H.~F.~Zhang,
  {\it {QCD corrections to the color-singlet $J/\psi$ production in deeply inelastic scattering at HERA}},
  {\em Phys.\ Rev.} {\bf D96} (2017), 091502, [\href{https://arxiv.org/abs/1705.05337}{{\tt arXiv:1705.05337}}].

\bibitem{Jpsi101}
  N.~Brambilla {\it et al.} [Quarkonium Working Group],
  {\it {Heavy quarkonium physics}}, [\href{https://arxiv.org/abs/hep-ph/0412158}{{\tt hep-ex/0412158}}].

\bibitem{Jpsi11}
  N.~Brambilla {\it et al.},
  {\it {Heavy Quarkonium: Progress, Puzzles, and Opportunities}},
  {\em Eur.\ Phys.\ J.} {\bf C71} (2011) 1534, [\href{https://arxiv.org/abs/1010.5827}{{\tt arXiv:1010.5827}}].

\bibitem{Jpsi12}
  J.~P.~Lansberg,
  {\it {New Observables in Inclusive Production of Quarkonia}}, [\href{https://arxiv.org/abs/1903.09185}{{\tt arXiv:1903.09185}}].

\bibitem{Jpsi13}
  H.~F.~Zhang, W.~L.~Sang and Y.~P.~Yan,
  {\it {Statistical analysis of the azimuthal asymmetry in the $J/\psi$ leptoproduction in unpolarized $ep$ collisions}},
  {\em JHEP} {\bf 1910} (2019) 234, [\href{https://arxiv.org/abs/1908.02521}{{\tt arXiv:1908.02521}}].

\bibitem{DISexp1}
  {\bf H1} Collaboration, C.~Adloff {\it et al.},
  {\it {Charmonium production in deep inelastic scattering at HERA}},
  {\em Eur.\ Phys.\ J.} {\bf C10} (1999) 373, [\href{https://arxiv.org/abs/hep-ph/9903008}{{\tt hep-ex/9903008}}].

\bibitem{DISexp2}
  {\bf H1} Collaboration, C.~Adloff {\it et al.},
  {\it {``Inelastic leptoproduction of $J/\psi$ mesons at HERA}},
  {\em Eur.\ Phys.\ J.} {\bf C25} (2002) 41, [\href{https://arxiv.org/abs/hep-ph/0205065}{{\tt hep-ex/0205065}}].

\bibitem{DISexp3}
  {\bf ZEUS} Collaboration, S.~Chekanov {\it et al.},
  {\it {Measurement of inelastic $J/\psi$ production in deep inelastic scattering at HERA}},
  {\em Eur.\ Phys.\ J.} {\bf C44} (2005) 13, [\href{https://arxiv.org/abs/hep-ph/0505008}{{\tt hep-ex/0505008}}].

\bibitem{DISexp4}
  {\bf H1} Collaboration, F.~D.~Aaron {\it et al.},
  {\it {Inelastic Production of $J/\psi$ Mesons in Photoproduction and Deep Inelastic Scattering at HERA}},
  {\em Eur.\ Phys.\ J.} {\bf C68} (2010) 401, [\href{https://arxiv.org/abs/1002.0234}{{\tt arXiv:1002.0234}}].

\bibitem{Jpsihadron1}
  P.~Artoisenet, J.~M.~Campbell, J.~P.~Lansberg, F.~Maltoni and F.~Tramontano,
  {\it {$\Upsilon$ Production at Fermilab Tevatron and LHC Energies}},
  {\em Phys.\ Rev.\ Lett.} {\bf 101} (2008) 152001, [\href{https://arxiv.org/abs/0806.3282}{{\tt arXiv:0806.3282}}].

\bibitem{Jpsihadron2}
  B.~Gong and J.~X.~Wang,
  {\it {Next-to-leading-order QCD corrections to $J/\psi$ polarization at Tevatron and Large-Hadron-Collider energies}},''
  {\em Phys.\ Rev.\ Lett.} {\bf 100} (2008) 232001, [\href{https://arxiv.org/abs/0802.3727}{{\tt arXiv:0802.3727}}].

\bibitem{Jpsihadron3}
  J.~P.~Lansberg,
  {\it {On the mechanisms of heavy-quarkonium hadroproduction}},''
  {\em Eur.\ Phys.\ J.} {\bf C61} (2009) 693, [\href{https://arxiv.org/abs/0811.4005}{{\tt arXiv:0811.4005}}].

\bibitem{Jpsiphoto1}
  M.~Krämer,
  {\it {QCD corrections to inelastic $J/\psi$ photoproduction}},
  {\em Nucl.\ Phys.} {\bf B459} (1996) 3, [\href{https://arxiv.org/abs/hep-ph/9508409}{{\tt hep-ex/9508409}}].

  \bibitem{Jpsiphoto2}
  P.~Artoisenet, J.~M.~Campbell, F.~Maltoni and F.~Tramontano,
  {\it {J/psi production at HERA}},
  {\em Phys.\ Rev.\ Lett.} {\bf 102} (2009) 142001, [\href{https://arxiv.org/abs/0901.4352}{{\tt arXiv:0901.4352}}].

\bibitem{Jpsiphoto3}
  C.~H.~Chang, R.~Li and J.~X.~Wang,
  {\it {$J/\psi$ polarization in photo-production up-to the next-to-leading order of QCD}},
  {\em Phys.\ Rev.} {\bf D80} (2009) 034020, [\href{https://arxiv.org/abs/0901.4749}{{\tt arXiv:0901.4749}}].

\bibitem{LHeC1}
  {\bf LHeC} Study Group, J.~L.~Abelleira Fernandez {\it et al.},
  {\it {A Large Hadron Electron Collider at CERN: Report on the Physics and Design Concepts for Machine and Detector}},
  {\em J.\ Phys.\ G.} {\bf 39} (2012) 075001, [\href{https://arxiv.org/abs/1206.2913}{{\tt arXiv:1206.2913}}].

\bibitem{LHeC2}
  {\bf LHeC} Study Group, J.~L.~Abelleira Fernandez {\it et al.},
  {\it {A Large Hadron Electron Collider at CERN}}, [\href{https://arxiv.org/abs/1211.4831}{{\tt arXiv:1211.4831}}].

\bibitem{PDG}
Particle Data Group, M.~Tanabashi {\it et al.},
  {\it {Review of Particle Physics}},
  {\em Phys.\ Rev.} {\bf D98} (2018),  030001.

\bibitem{NRQCD}
  G.~T.~Bodwin, E.~Braaten and G.~P.~Lepage,
  {\it {``Rigorous QCD analysis of inclusive annihilation and production of heavy quarkonium}},
  {\em Phys.\ Rev.} {\bf D51} (1995) 1125, Erratum: [{\em Phys.\ Rev} {\bf D55} (1997) 5853], [\href{https://arxiv.org/abs/hep-ph/9407339}{{\tt hep-ex/9407339}}].

\bibitem{Berezhnoy:2020aox}
  A.~Berezhnoy, I.~Belov and A.~Likhoded,
  {\it {The production of excited states of doubly heavy baryons at the Large Hadron Collider}},
  [\href{https://arxiv.org/abs/2005.04760}{{\tt arXiv:2005.04760}}].

\bibitem{Sjostrand:2006za}
  T.~Sjostrand, S.~Mrenna and P.~Z.~Skands,
  {\it {PYTHIA 6.4 Physics and Manual}},
  {\em JHEP} {\bf 0605} (2006) 026, [\href{https://arxiv.org/abs/hep-ph/0603175}{{\tt hep-ex/0603175}}].

\bibitem{fragmentation}
  C.~Peterson, D.~Schlatter, I.~Schmitt and P.~M.~Zerwas,
  {\it {Scaling Violations in Inclusive e+ e- Annihilation Spectra}},
  {\em Phys.\ Rev.} {\bf D27} (1983) 105.

\bibitem{projector}
  A.~Petrelli, M.~Cacciari, M.~Greco, F.~Maltoni and M.~L.~Mangano,
  {\it {NLO production and decay of quarkonium}},
  {\em Nucl.\ Phys.} {\bf B514} (1998) 245, [\href{https://arxiv.org/abs/hep-ph/9707223}{{\tt hep-ex/9707223}}].

\bibitem{Yu:2017zst}
  F.~S.~Yu, H.~Y.~Jiang, R.~H.~Li, C.~D.~Lu, W.~Wang and Z.~X.~Zhao,
  {\it {Discovery Potentials of Doubly Charmed Baryons}},
  {\em Chin.\ Phys.} {\bf C42} (2018), 051001, [\href{https://arxiv.org/abs/1703.09086}{{\tt arXiv:1703.09086}}].

\bibitem{EIC}
A.~Accardi  {\it et al.}
  {\it {Electron Ion Collider: The Next QCD Frontier}},
  {\em Eur.\ Phys.\ J.} {\bf A52} (2016), 268, [\href{https://arxiv.org/abs/1212.1701}{{\tt arXiv:1212.1701}}].

\end{thebibliography}
\end{document}